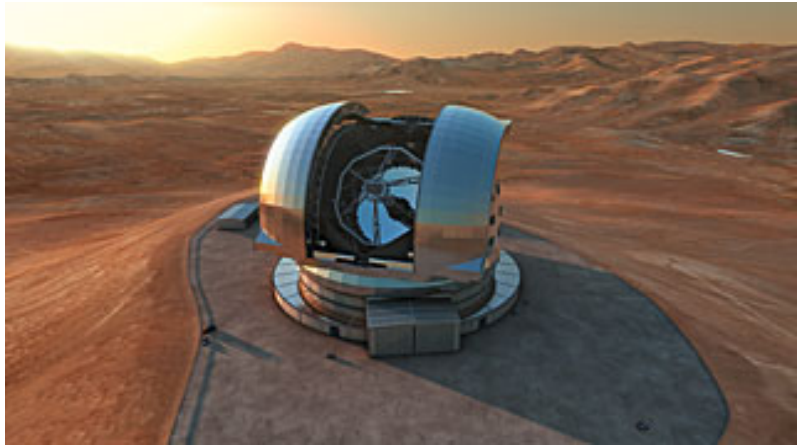

| Document Title | ELT-MOS White Paper: Science Overview & Requirements |
|---|---|
| Issue | 1.0 |
| Date | 22 February 2013 |
| Editors | Chris Evans (UK ATC) & Mathieu Puech (GEPI) |

Contributors:
Beatriz Barbuy, Nate Bastian, Piercarlo Bonifacio, Elisabetta Caffau, Jean-Gabriel Cuby, Gavin Dalton, Ben Davies, Jim Dunlop, Chris Evans, Hector Flores, Francois Hammer, Lex Kaper, Bertrand Lemasle, Simon Morris, Laura Pentericci, Patrick Petitjean, Mathieu Puech, Daniel Schaerer, Eduardo Telles, Niraj Welikala, Bodo Ziegler

# EXECUTIVE SUMMARY

The workhorse instruments of the 8-10m class observatories have become their multi-object spectrographs (MOS), providing comprehensive follow-up to both ground-based and space-borne imaging. With the advent of deeper imaging surveys from, e.g., the *HST* and VISTA, there are a plethora of spectroscopic targets which are already beyond the sensitivity limits of current facilities.

This wealth of targets will grow even more rapidly in the coming years, e.g., after the completion of ALMA, the launch of the *JWST* and *Euclid*, and the advent of the LSST. Thus, one of the key requirements underlying plans for the next generation of ground-based telescopes, the Extremely Large Telescopes (ELTs), is for even greater sensitivity for optical and infrared (IR) spectroscopy. Moreover, with only three ELTs planned, the need to make efficient use of their focal planes becomes even more compelling than at current facilities given the large construction and operational costs.

Here we revisit the scientific motivation for a MOS capability on the E-ELT, combining updated elements of science cases advanced from the Phase A instrument studies, with new science cases which draw on the latest results and discoveries. These science cases address key questions related to galaxy evolution over cosmic time, from studies of resolved stellar populations in nearby galaxies out to observations of the most distant galaxies, and are used to identify the top-level requirements on an 'E-ELT/MOS'.

We argue that several of the most compelling ELT science cases demand MOS observations, in highly competitive areas of modern astronomy. Recent technical studies have demonstrated that important issues related to e.g. sky subtraction and multi-object AO can be solved, making fast-track development of a MOS instrument feasible. To ensure that ESO retains world leadership in exploring the most distant objects in the Universe, galaxy evolution and stellar populations, we are convinced that a MOS should have high priority in the instrumentation plan for the E-ELT.

# TABLE OF CONTENTS

# 1   INTRODUCTION

In parallel to the Phase B design of the European Extremely Large Telescope (E-ELT; e.g. Gilmozzi & Spyromilio, 2008), nine Phase A instrument studies were undertaken (see Ramsay et al. 2010). These studies spanned a vast range of parameter space, in part to evaluate the relative merits of different capabilities toward the scientific cases advanced for the E-ELT, while also exploring the instrument requirements and technology readiness of likely components. Three of the Phase A studies were of MOS instruments: EAGLE (Cuby et al. 2010), OPTIMOS-EVE (Navarro et al. 2010), and OPTIMOS-DIORAMAS (Le Fèvre et al. 2010). Each explored different parameter space in terms of the image quality provided by adaptive optics (AO), number of targets, spectral coverage, spectral resolution, and imaging capability.

The ESO instrument roadmap has identified two first-light instruments for the E-ELT (Ramsay et al. 2012): a near-IR imager and a (red-)optical/near-IR integral field unit (IFU) spectrograph, with requirements similar to the Phase A studies for MICADO (R. Davies et al. 2010) and HARMONI (Thatte et al. 2010), respectively. These will exploit the limits of image quality of the E-ELT via high performance AO, but will necessarily be limited in their spatial extent on the sky: i.e. a 1 to 2 arcmin field-of-view imager, and a monolithic (i.e. single-target) IFU. A HARMONI-like instrument will be well suited to spectroscopy of individual high-$z$ galaxies, and stars in very dense regions (e.g. inner parts of spirals, cluster complexes), but the larger samples needed to explore galaxy evolution – at high and low redshifts – will require MOS observations. In the following sections we give an overview of illustrative science cases for an 'ELT MOS', followed by consideration of the instrument requirements to enable the proposed observations.

The E-ELT has been designed to have excellent image quality over a 10 arcmin diameter field-of-view, but the infrastructure required for AO guide stars and other sensors limits it to an effective field of ~7 arcmin. In this document we therefore assume a patrol field with an equivalent diameter of 7 arcmin, offering the attractive prospect of a ~40 arcmin$^2$ field on a 40m class telescope for spectroscopy.

We note that the E-ELT has been designed with AO integrated into the telescope, with a large adaptive mirror ('M4') and a fast tip-tilt mirror ('M5') to correct for the turbulence in the lower layers of the atmosphere via ground-layer adaptive optics (GLAO). This is envisaged as the basic level of image quality for E-ELT observations; higher-performance AO will be delivered by dedicated modules, or within the instruments themselves. The level of image quality (i.e. AO correction) required for each case is discussed in the following sections; these are divided into one of two types: 'high definition', where high-performance AO is required for tens of objects, and 'high multiplex', where GLAO performance is sufficient for 100s of objects.

# 2   SC1: 'FIRST LIGHT' – SPECTROSCOPY OF THE MOST DISTANT GALAXIES

## 2.1   Probing the epoch of reionisation

Some 380,000 years after the Big Bang the temperature of the Universe was low enough that the hydrogen-dominated intergalactic medium (IGM) which pervades the Universe became neutral. The IGM today is fully ionised, heated by the integrated ultraviolet (UV) emission from galaxies and active galactic nuclei (AGN). However, how and when the IGM turned from neutral to fully ionised is a matter of great debate.

Observations of the highest-redshift quasars indicate the transition to a fully-ionised IGM occurred no later than ~1 Gyr after the Big Bang, by $z$~6 (e.g. Fan et al. 2006). Quasar spectra can be used to measure the evolution of the Ly-$\alpha$ and Ly-$\beta$ effective optical depth with redshift and at $z$~6-6.5 they show a complete 'Gunn-Peterson trough', which is sensitive to very small amounts of neutral hydrogen (neutral fraction $x_{HI}$ < $10^{-3}$), thus constraining the end of reionisation. Another constraint comes from measurements of the electron-scattering optical depth from the polarisation of the cosmic-microwave background, which has demonstrated the presence of ionising sources in the early Universe at $z$~10-15 (Spergel et al. 2007). The latest results from *WMAP* (Hinshaw et al. 2013) indicate that the IGM may have been half-ionised by $z$~10, some 500 Myr earlier. The reionisation history between these limits remains basically unknown and is difficult to trace with present-day instruments. Whether the reionisation occurred slowly over this period, or if there were several sporadic periods of ionisation from more than one generation of ionising sources, is completely unknown.

The main sources of reionisation have remained elusive so far, probably due to their faintness (e.g. Choudhury & Ferrara 2007; Bunker et al. 2009; Robertson et al. 2010; Bouwens et al. 2012). The quest for these sources, which produced the UV radiation field that reionised the IGM, is therefore intimately related to the search for the first, most distant galaxies (Ellis et al. 2013; McLure et al 2013; Robertson et al. 2013).

*Ultra-deep ELT-MOS observations of faint continuum-selected sources (from visible and near-IR imaging) will provide the necessary observational constraints to robustly determine the Ly-$\alpha$ properties and hence the ionisation state of the IGM from redshift 5 to 13 (Sect. 2.2). Such a programme will also determine the properties of these first, bright galaxies including their ISM, outflows and stellar populations (Sect. 2.3).*

*Such a programme will ideally complement JWST observations. JWST will provide spectral energy distributions (SEDs) of exquisite quality from the near IR to the mid IR (20 $\mu$m), and a more homogeneous spectral coverage than on the ground. However, its sensitivity in spectroscopy will be significantly less than that of the E-ELT, particularly for continuum spectroscopy.*

## 2.2  Looking for the low-luminosity sources responsible for the reionisation

The evolution of the Ly-$\alpha$ emitter luminosity function, LF(Ly-$\alpha$), with redshift can be used to derive constraints on the volume-weighted neutral-hydrogen fraction in the IGM (e.g. Malhotra & Rhoads 2004; Dijkstra et al. 2007). The LF(Ly-$\alpha$) method is sensitive to a larger neutral-hydrogen fraction than the Gunn-Peterson test and is therefore one of the most promising methods to constrain the reionisation history of the Universe. However, current estimates are subject to considerable uncertainties and, most importantly, samples are limited to below $z\sim7$ (Clément et al. 2012). Among the main uncertainties are the very small samples of Ly-$\alpha$ emitters, the small fraction of spectroscopic confirmations, and sparse knowledge of source properties such as their star-formation rates (SFRs), SF histories, ages, outflow properties, etc.

An alternative approach is to target continuum-selected Lyman-break Galaxies (LBGs) and measure the equivalent width of their Ly-$\alpha$ emission, the fraction of them showing emission, and other related quantities. Such measurements, together with other methods (e.g. Hayes et al. 2011), indicate that beyond *z*~7 we are starting to probe an epoch at which the IGM is increasingly neutral. For instance, attempts to detect Ly-α at higher redshifts, using many tens/hundreds of hours on 8-10m class telescopes, have been less successful (e.g., Pentericci et al. 2011; Schenker et al. 2012), and the current spectroscopic redshift record is 'only' $z = 7.231$ (Ono et al. 2012). There is no indication that the physical properties of these galaxies change quickly at these redshifts, so it is plausible to infer that the reduced visibility is due to an increased amount of neutral hydrogen in their vicinity. Depending on the model, the neutral hydrogen required matching the observations ranges from 20 to 60% (e.g., Dijkstra, Mesinger & Wyithe 2011; Jensen et al. 2013).

The current observations, combined with disappointing returns from narrow-band Ly-α searches at *z*>7 (see review by Dunlop 2012), show that significant improvements are required to probe the increasingly neutral Universe at *z*>7. Ultra-deep *Y*- to *H*-band E-ELT spectroscopy of large samples of faint continuum-selected sources is required to determine their Ly-$\alpha$ properties and, hence, the ionisation state of the IGM from *z*~7-15. These objects will be known from visible/near-IR imaging with facilities such as the VLT, *HST*/WFC3, and *JWST.* Specifically, with an ELT MOS we aim to (see Fig. 1):

- Measure Ly-$\alpha$ line fluxes up to ~40 times deeper than current samples at *z*~7 in blank fields, ~40 times deeper than current searches for Ly-$\alpha$ emission from *z*~7-8.5 lensed galaxies, and >100 times deeper than the current searches at *z*~8-10 (Stark et al. 2007);
- Extend Ly-$\alpha$ searches up to *z*~13, and undertake spectroscopy of large numbers of galaxies down to $m_{AB}$~30, which are presumably the dominant population of reionisation sources, unachievable with other planned E-ELT instruments and perfectly suited to the expected source density;
- Reach an effective Ly-$\alpha$ transmission ($f_{eff}$) of 10% for very faint objects and correspondingly lower values for brighter objects. Even 1 (10)% transmission will be reachable for objects with $m_{AB}$ = 30 (32) for sources behind lensing clusters, benefiting from gravitational magnification by a factor of 10.

Taken together this unique information will allow us to accurately determine the hydrogen ionisation fraction in the IGM and its evolution with redshift, thus measuring the cosmic reionisation history and determining important properties of the first galaxies in the Universe.

Typical Ly-$\alpha$ transmissions are expected to be of the order of ~20% at $z$~7 and between 5-40% at $z$~8 for galaxies with $M_{UV}$ = −18 to −22 (i.e. $m_{AB}$ = 29 to 25), from the simulations of Dayal & Ferrara (2012). These models, accounting for clustering effects, predict a higher Ly-$\alpha$ transmission for brighter (more massive) sources. They are compatible with the current Ly-$\alpha$ fraction measured spectroscopically in LBGs at redshifts $z$~6-7, from the latest VLT and Keck campaigns (e.g. Pentericci et al. 2011; Schenker et al. 2012). Measuring the Ly-$\alpha$ fluxes and equivalent widths of a large sample of galaxies (e.g. more than 50 galaxies with $m_{AB}$ = 28 down to fluxes of $2 \times 10^{-18}$ erg s$^{-1}$cm$^{-2}$) allows one already to distinguish different reionisation scenarios – e.g. patchy versus smooth reionisation – as shown by Treu et al. (2012). Fluxes fainter by a factor of 10-20 will be reached by an ELT-MOS, meaning that more stringent tests of different reionisation models will be possible with the E-ELT. The corresponding source density, e.g. 2-5 arcmin$^{-2}$ mag$^{-1}$ at $m_{AB}$ = 29 for LBGs at $z$~8 (see Fig 1b) translates to 80-200 sources in the E-ELT patrol field. An ELT-MOS will ideally complement *JWST*-NIRSpec observations of the brightest objects at low spectral resolution.

For objects where Ly-α will remain elusive even with the E-ELT, and in addition to Ly-α, other emission lines are expected from these distant galaxies. Indeed, there are encouraging signs that any demise in observable Ly-α emission may be compensated by increasingly bright, high-ionisation UV lines, such as CIII] at $\lambda_{rest}$ = 1909Å. Support for this comes from observations of extreme [OIII] emission in some low-luminosity, (likely) low metallicity, star-forming galaxies at $z$~2 (van der Wel et al. 2011), as well as indications from broad-band photometry that such strong high-ionisation emission may become increasingly prevalent at high-$z$ (e.g. de Barros et al. 2012; Stark et al. 2012; Labbé et al. 2012) due to higher specific star-formation rates and/or to harder UV emission from massive stars with significantly subsolar metallicities.

The CIII] line is expected to have a rest-frame $EW_{CIII]}$ ~ 0.15 x $EW_{Ly-\alpha}$. Thus, at $z$~7.5, CIII] line fluxes of ~$3 \times 10^{-18}$ and $3 \times 10^{-19}$ erg s$^{-1}$cm$^{-2}$ are expected from galaxies with $H_{AB}$ = 27.5 and 30.0 mag, respectively. If CIII] transpires to be a key emission-line for redshift determination at $z$>7, then the importance of following it out to $z$~10 (where several LBGs have now been detected - Zheng et al. 2012; Coe et al. 2013; Ellis et al. 2013) provides an argument for extending near-IR spectroscopy into the *K*-band. Lastly, we note that the goal of detecting HeII ($\lambda_{rest}$ = 1640Å) remains vital in establishing when high-$z$ galaxies start to be dominated by light from massive, extremely metal-poor ('Pop III') stars.

## 2.2 Probing the physical properties of the 'first-light' galaxies

For the brightest sub-population of these first galaxies, an ELT-MOS observing programme will also determine the physical properties of there ISM, stellar populations and, if present, outflows. In addition to detecting emission-lines in such distant galaxies, an ELT-MOS will have a sufficiently large patrol field to follow-up the near-IR imaging surveys which will be used to identify LBGs within the reionisation epoch.

ELT-MOS observations extending into the *H*-band will provide us with the ability to detect the Lyman-break ($\lambda_{rest}$ = 912Å) out to $z$~17, and the rest-frame UV absorption lines at $\lambda_{rest}$ = $\lambda\lambda$1200-2000Å out to $z$~10; this region includes prominent spectral features from Ly-α $\lambda$1215, NV $\lambda$1240, HeII $\lambda$1640, CIV $\lambda$1549, SiIV $\lambda\lambda$1393,1402, SiII $\lambda$1260, OI $\lambda$1303, and CIII] $\lambda$1908. As noted above, these features may offer redshift determinations even if Ly-α is completely absorbed, as well as providing important information on the ISM, large-scale outflows, metal enrichment, and the IMF.

Low-resolution spectroscopy can be sufficient to derive redshifts, but moderate resolution ($R$~5,000) is required to study the detailed physics of galaxies, as derived from the line strengths and profiles of the ISM lines (such as SiIV, OI and CIV) and stellar absorption lines (NV, CIV, SiIV, NIV, with some ISM contribution in these). Furthermore, the identification of significant velocity offsets between UV ISM lines, Ly-α, and photospheric lines in galaxies at $z$~3 led to the discovery that a large fraction of star-forming galaxies at $z$~3 are driving strong winds – 'superwinds' – a feedback process thought to be the dominant mechanism which expels baryons from galaxies at these early times (e.g., Wilman et al. 2005). The likely shallower potential wells of typical galaxies at $z$>7 suggests that outflows from high-$z$ galaxies could have cleared material from the galaxy and surroundings, allowing ionising photons to escape, as well as enriching the IGM with metals. This will shed light on the problem of how the IGM was reionised and enriched.

Rest-frame UV spectroscopy of this type was pioneered in the 1990s using the Keck telescope to observe galaxies at $z\sim3$ (Shapley et al. 2003) and progressively extended up to $z\sim4$-5 using Keck and the VLT (Vanzella et al. 2005, 2009; Douglas et al. 2010; Jones et al. 2011). This will likely continue up to $z\sim6$ over the coming decade with new instruments such as VLT-KMOS and Keck-MOSFIRE but continuum and absorption-line studies in the UV at $z>7$ will remain out of reach for the current generation of 8-m class telescopes and of JWST. In other words, this rest-frame UV continuum work can only be done efficiently with a wide-field, multi-object capability on an ELT.

Large samples over large volumes are essential to smooth over cosmic variance, and to test the large-scale structure of galaxies and the inhomogeneity of the IGM as the Universe is reionised. We therefore need:

- A multi-object capability to mitigate the relatively long integration times necessary to detect continuum and absorption lines in these objects down to $H_{AB}\sim28$;
- An IFU capability to allow us to resolve their clumpy internal structures (e.g., Law et al. 2009);
- Good image quality because $z>7$ sources are compact, with half-light radii <0.15 arcsec (Grazian et al. 2012; Ono et al. 2013).

Interestingly, the use of strong lensing provided by massive galaxy clusters to probe instruments to the faintest possible limits is now well established (see e.g. Smail et al. 1997; Richard et al. 2006, 2008; Stark et al. 2007). Gravitational lensing can provide access to exceptionally faint objects (~2-3 mag below the normal detection limits), but their number per field will be relatively small. For example, with a Ly-$\alpha$ sensitivity of ~$10^{-19}$ erg s$^{-1}$ cm$^{-2}$, an ELT-MOS will be able to determine an effective Ly-$\alpha$ transmission down to ~0.1 out to $z\sim9$ for objects with $m_{AB}(UV) = 32$, and magnification by ~2 mag. In addition to the magnification, the image of the background galaxy is stretched, so it is possible to study the spatially-resolved ISM physics in detail, which is impossible to do without lensing (e.g., Swinbank et al. 2004; Combes et al. 2012).

Finally, spatially-resolved spectroscopy of the brightest sub-sample of the first galaxies will enable us to probe the geometry and sizes of the outflows and infer the mass-outflow rates. Although the high-$z$ galaxies are small, their faintness means that diffraction-limited spectroscopy is not required; the basic requirement is to reach sufficient angular resolution to concentrate the light into 50-100 mas scales with ~10 angular resolution elements across the target. Such sampling will allow us to probe the spatially-resolved stellar populations within the galaxy, and between multiple components within the same system. Many of the LBGs detected at $z$=5–6 show multiple components or companions of 1-2 arcsec scales in projection (e.g., Conselice & Arnold 2009). If this trend continues to higher-$z$ (as expected in a hierarchical mass-assembly model, and also based on results on lensed objects, e.g., Swinbank et al. 2009), then spatially resolving the velocity offsets and stellar populations within multiple components/merging systems provides an important way of constraining the dynamical masses and stellar mass-to-light ratios of these galaxies.

## 2.3  Finding the targets – synergies with other facilities

In the coming years, current or upcoming near-IR faciliites (Keck-MOSFIRE, VLT-KMOS, VLT-MOONS) will be used to explore targets beyond $z\sim7$. However, if the observed decrease in Ly-α emission is confirmed, such work will become increasingly hard with 8-10m class telescopes. Moreover, the latest HST/WFC3 observations are already revealing sources that are clearly beyond the capabilities of current ground-based facilities. Thus, most of the current and future candidates – probably in the 100s – are likely to remain of interest when the E-ELT arrives. Ahead of this, work will be required to extend the existing ultra-deep imaging to a larger number of fields on the sky, e.g., data taken for the HST Frontiers Fields, near gravitational lenses.

The JWST is often presented as the 'first-light machine', but it is important to note that while it will have superb imaging sensitivity (offering a vast discovery space), it will be limited in terms of its spectroscopic sensitivity due to its (relatively) small collecting area and its readout noise limited performance. An ELT-MOS will thus provide the capability to obtain spectra of most of the targets discovered/identified with the JWST, down to the faintest magnitudes.

High resolution ($\delta z<\sim 0.001$) and high S/N Ly-$\alpha$ observations, coupled with deep ALMA observations of the [CII] 158µm emission line and far-IR continuum, will allow us to address a number of interesting issues. In particular, the ALMA observations will allow us to determine the systemic redshift of the galaxies with unprecedented accuracy. Once the precise redshifts of the galaxies are known, the asymmetric profile of the Ly-$\alpha$ lines can be accurately modelled in terms of IGM absorption and outflows (e.g., Dayal et al. 2011), thus tightly constraining the neutral-hydrogen fraction (without being affected by uncertainties associated with giant HII regions surrounding luminous quasars at similar redshifts).

Further into the future, one can foresee valuable observing programmes using the Square Kilometre Array (SKA) to measure HI 21 cm gas properties of galaxies with extant E-ELT and ALMA data, and also combinations of the above with future X-ray missions (e.g., the *International X-ray Observatory*).

## 2.4 Requirements

*Spectral Resolution :* $R>3,000$ is required to target emission lines between the OH sky lines (e.g., Navarro et al. 2010; Villanueva et al. 2012). Resolving Ly-$\alpha$ profiles with typical FWHM ~150-500 km/s (Tapken et al. 2007, Kashikawa et al. 2006) requires at least $R=4,000$ to marginally sample the narrowest lines, with the optimal requirement being $R=5,000$ to separate the effect of the IGM transmission (i.e. the ionisation state of the IGM) from radiation transfer effects in the host galaxy and its close environment, such as galactic winds, known to alter the Ly-$\alpha$ line profile (e.g., Santos 2004; Verhamme et al. 2006).

*Multiplex/Observing modes :* Current Ly-α observations reach depths of ~3-10×$10^{-18}$ erg/s/cm$^2$ (e.g., Pentericci et al. 2011; Schenker et al. 2012). Ly-α is expected to have fluxes as faint as ~$10^{-18}$ erg s$^{-1}$ cm$^{-2}$ at $z\sim7.5$ in galaxies with $H_{AB}$ = 30.0 mag, but if the possible decrease of Ly-α emission at $z>7$ is confirmed (Pentericci et al. 2011; Ono et al. 2012), the required flux depth for $m_{AB}=28-30$ sources would be closer to ~$10^{-19}$ erg s$^{-1}$ cm$^{-2}$. An ELT-MOS should be able to obtain a S/N~10 (per resolution element) for such fluxes in a reasonable integration times (a few tens of hours) for targets down to $H_{AB} \sim 29$ (e.g. Fig. 1; left panel).

Considering current observations with the *HST*/WFC3 (e.g., Bouwens et al. 2011, Grazian et al. 2012, Oesch et al. 2012; Ellis et al. 2013; McLure et al. 2013) and integrating the sources with $H_{AB} \leq 30$ over $6.5 \leq z \leq 9.5$, the surface density of LBGs is found to be on average ~10 arcmin$^{-2}$ (Fig. 1; right panel); i.e., in this redshift range the number of potential targets for emission line detection within a 7(10) arcmin diameter patrol field is ~300(800) galaxies. At $R>3,000$, at least half of the *J* and *H*-bands are free of OH sky lines (e.g., Navarro et al. 2010; Villanueva et al. 2012), which lead to potential target densities of at least ~150 (400) galaxies in emission (single-line detection). Both AO-corrected and integrated-light (i.e. GLAO) observations of candidates will advance this field; via excellent sensitivity in the case of the former, and a large multiplex in the latter case.

Detections in the continuum or interstellar absorption-lines will be limited to the brightest objects. Simulated observations suggest that such a limit is $m_{AB}\sim27$ (see Fig. 2, and perhaps a magnitude deeper for 'Large Programme'-like deep observations). The surface density at these magnitudes is an order of magnitude smaller, which leads to ~15-40 targets in the E-ELT patrol field. Taking into account the spectroscopic success rate of metal lines in the rest-frame UV and extrapolating number counts from the *HST*, one indeed finds 40 sources at $z\sim7$ down to $J_{AB}\sim28$ and 20 galaxies down to $H_{AB}\sim28$ at $z\sim8$ (Evans et al. 2010). For such faint sources, spatially-resolved information is desirable, which calls for an AO-fed multi-IFU observing mode for an ELT-MOS.

*IFU spatial sampling:* In considering the likely structures in the earliest galaxies, it is important to first note that all imaging observations of very high-z galaxies undertaken to date suggest that they are extremely compact, with half-light radii <0.15 arcsec (Oesch et al. 2010; Grazian et al. 2012 ; Bouwens et al. 2012; Ono et al. 2013). However, these measurements are all based on *HST* imaging, which is inevitably most sensitive to high surface-brightness peaks in emission. Most galaxies at $z>2$ indeed appear clumpy (see also SC2). The typical clump size is ~1 kpc, which translates into ~180 mas above $z\sim6$ (with little evolution as a function of redshift at higher redshifts). If one wants to be able to resolve such internal structures, an IFU spatial sampling of at least 90 mas is therefore required. A spatial sampling of 40 mas would provide more than two spatial resolution elements, and hence more accurate measurements of the internal kinematics and other properties of clumps.

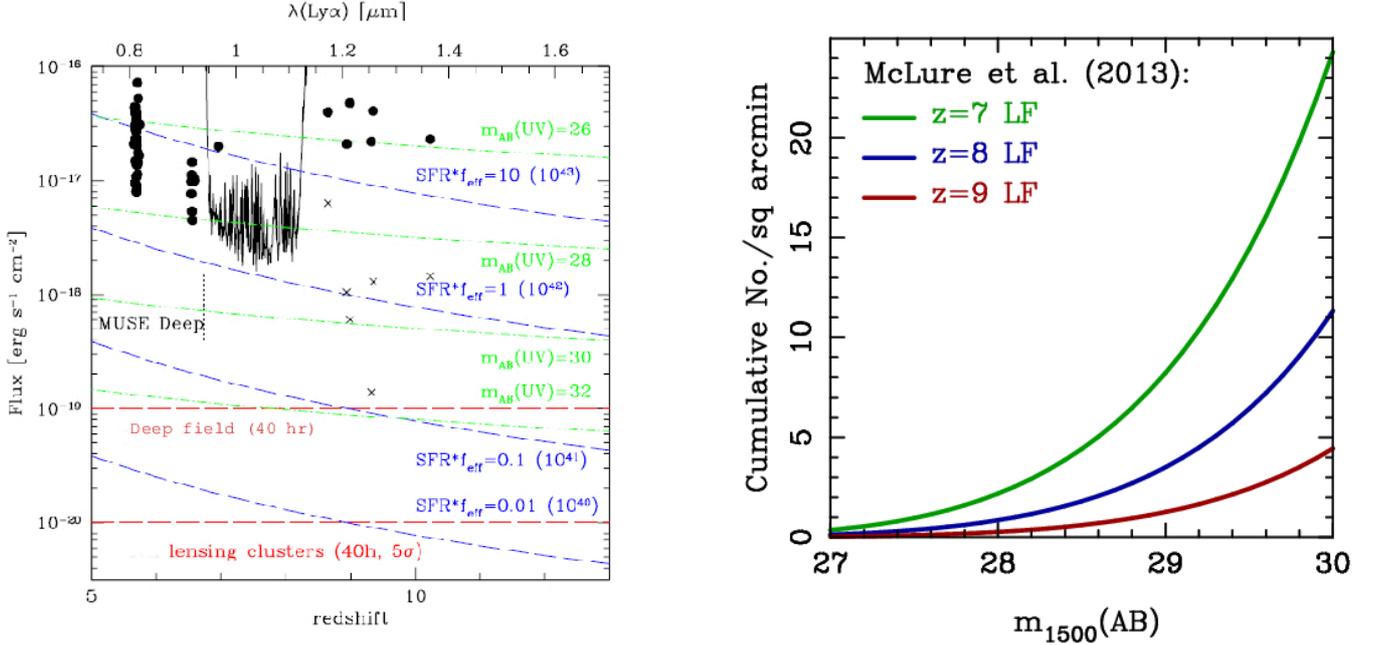

**Figure 1:** *Left panel:* Ly-α sensitivities for an ELT-MOS compared to expectations and the deepest Ly-α observations available at $z>5$. *Blue dashed lines:* Expected fluxes as a function of $z$ for a range of SFR×$f_{eff}$ from 10 to 0.01 $M_\odot$ yr$^{-1}$ corresponding to L(Ly-α) from $10^{43}$ to $10^{40}$ erg s$^{-1}$; $f_{eff}$ denotes the effective Ly-α transmission (including the IGM and other possible losses, e.g. inside the galaxy). *Green dash-dotted lines:* the expected Ly-α flux for star-forming galaxies with rest-frame UV continuum magnitudes as shown ($f_{eff}$ = 1). *Filled circles:* Observations from Shimasaku et al. (2006), Kashikawa et al. (2006), and Ota et al. (2008) at $z$ = 5.7, 6.5, and 7. C*rosses:* intrinsic fluxes for lensed candidates from Stark et al. (2007). *Black line:* flux limit from deep Keck-NIRSPEC spectroscopy from Richard et al. (2008). An ELT-MOS will provide a gain of a factor of ≥40 factor in sensitivity, extending up to z~13! For galaxies with $m_{AB}$ = 30 we will be able to measure IGM transmissions down to 10 % or even fainter in lensing clusters. *Right panel:* Observed integrated surface density of high-redshift star-forming galaxies per unit redshift interval at $z$ = 7, 8, and 9 as a function of observed AB near-IR (rest-frame UV) magnitude, as derived from the latest determination of the evolving high-redshift galaxy luminosity function from the UDF12 and CANDELS WFC3/IR *HST* programs (McLure et al. 2013).

To investigate the expected sensitivity in continuum/absorption line detections, we simulated spatially-resolved E-ELT observations of a clumpy galaxy at $z$~7 with $t_{int}$= 40 hr. We used the Shapley et al. (2003) LBG template to mimic real UV interstellar lines, while the morphology and kinematics were set-up using hydrodynamical simulations of rotating clumpy galaxies from Bournaud et al. (2008), which were rescaled to expected sizes and fluxes at $z$=7 (see Fig. 2). As shown in Fig. 2, the simulated observations encompass the range of observed magnitudes and sizes: 'compact' ($R_{half}$~100mas), 'average' ($R_{half}$~150mas), and 'large' ($R_{half}$=220 mas), thus sampling the observed distributions (see histograms of $J_{AB}$ and $R_{half}$ in the lower and right-hand panels of Fig. 2). We then used spectra in the datacubes with S/N >2 to construct the optimal composite spectrum in terms of S/N (see Fig. 3). Results show that S/N~3-5 can be reached for $J_{AB}$~27 targets within 40 hr of integration time depending on IFU sampling and target size.

In terms of S/N, the simulations suggest that optimum pixel scales are of the order of 140-250 mas, depending on the target size (see Figs. 2 & 3). However, this is incompatible with resolving the internal structures of these galaxies. Conversely, too small a pixel size (i.e., <40mas) results in an important loss in S/N in sources with the largest sizes due to readout-limited, non-optimal performance (e.g. Fig. 4). In summary, simulations suggest the optimal IFU spatial sampling should be in the range of 40-90 mas, depending on whether one wants to optimise the IFU pixel scale to spatially-resolve internal features such as clumps (closer to 40 mas/pixel), or to maximise the S/N of continuum/absorption line features (closer to 90 mas).

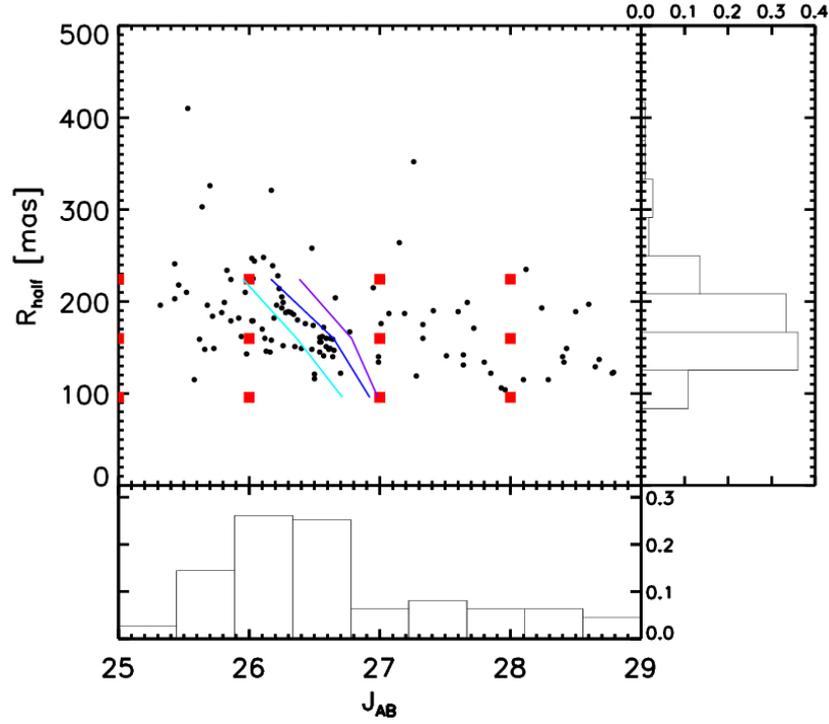

**Figure 2:** $R_{half}$ vs. $J_{AB}$ plane of observed LBG candidates at $z\sim7$ (black squares; Grazian et al. 2012) with the new simulated IFU observations indicated by the red squares. Each $J_{AB}/R_{half}$ combination was simulated 32 times with different noise realizations. For each simulated datacube, a S/N-optimal spectrum was constructed by stacking the IFU spaxels where S/N>2 until the S/N of the resulting integrated spectrum stops increasing (Rosales-Ortega et al. 2012), as shown in Fig. 3. The cyan, blue, and violet lines show the approx. detection limit for S/N (cont.) =5 for integrated spectra of sources with pixel scales of 40, 80, and 120 mas, respectively.

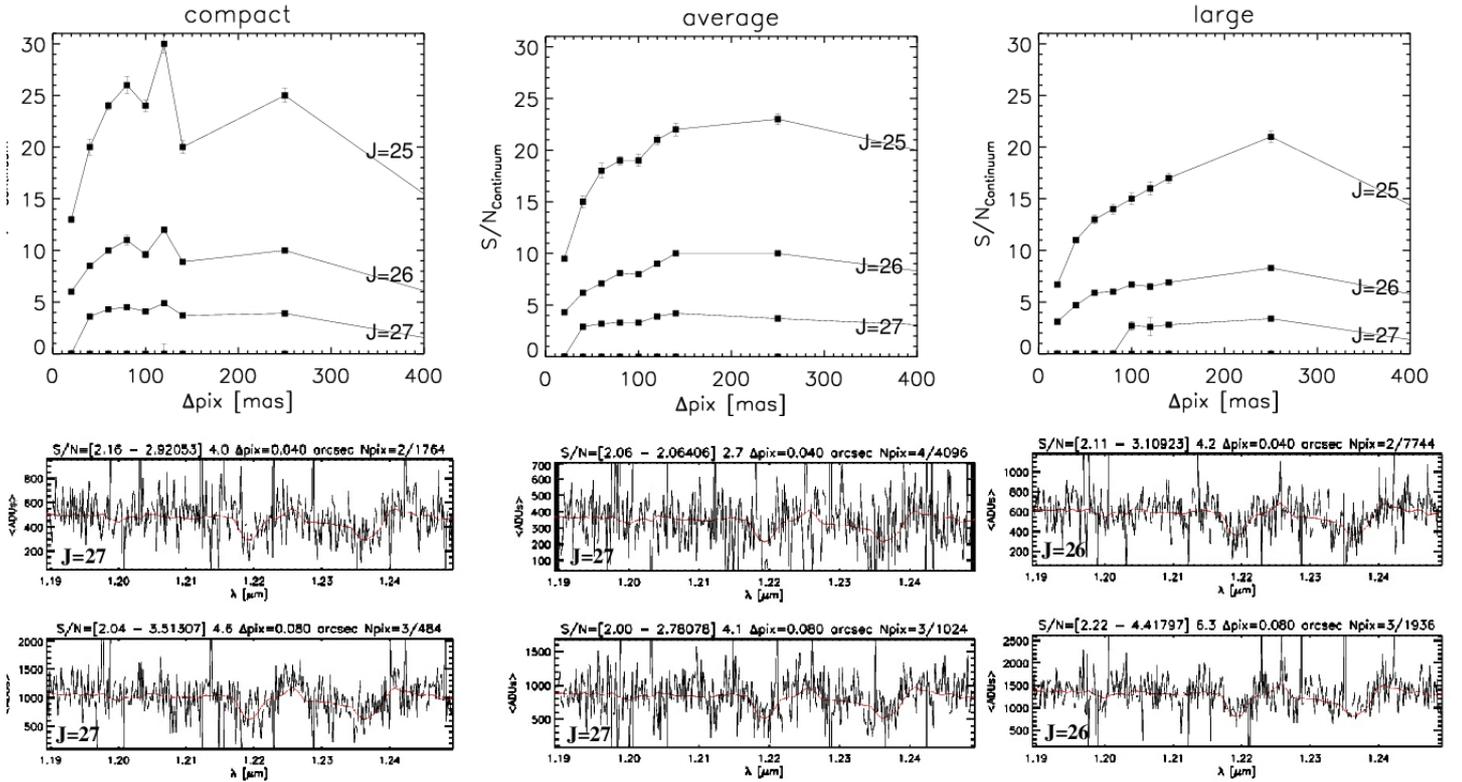

**Figure 3:** Simulated IFU observations of UV interstellar absorption lines redshifted to $z\sim7$ for a range a spatial sampling and object apparent magnitudes ($t_{int}$ = 40 hrs on source, dark-time conditions, MOAO correction with ensquared energy of 30 % in 80x80 mas²).
*Upper panels:* Average continuum S/N from integrated spectra of a sources at $z\sim7$ for the simulations of 'compact', 'average', and 'large' galaxies. *Middle & lower panels:* Examples of integrated spectra for $J_{AB}$=27 and 40 and 80mas/pixel, respectively. The range of continuum S/N for the spectra used to construct the combined spectrum are shown in parentheses at the upper left of each spectrum, followed by the S/N of the integrated spectrum itself. The spectra for the 'large' galaxies are for $J_{AB}$ = 26.

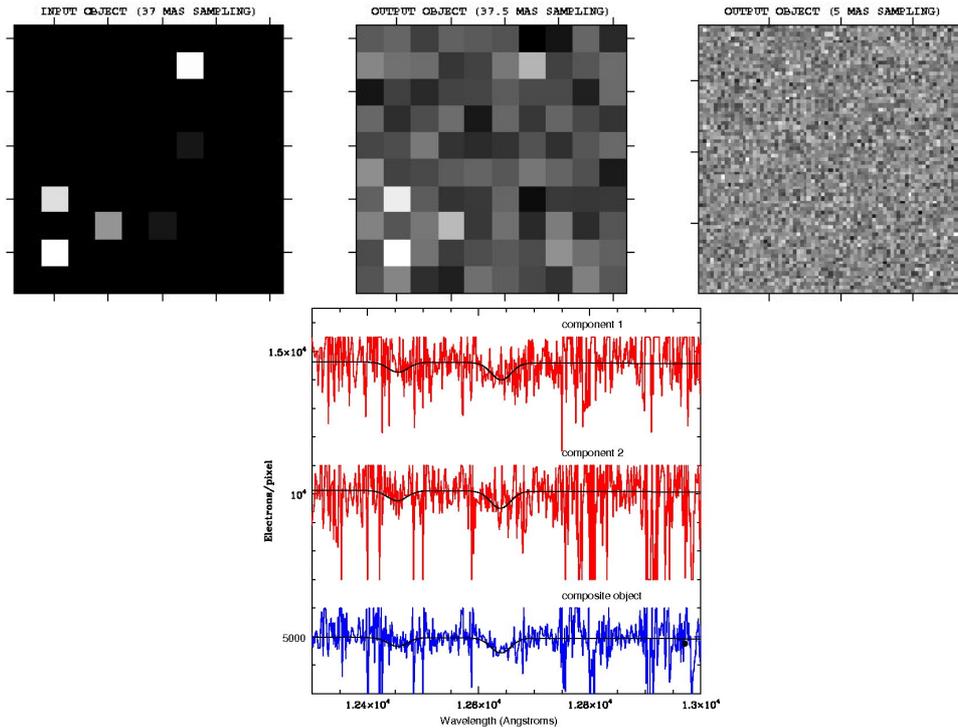

**Figure 4:** Simulations of a clumpy high-z galaxy at *z*~7 with $J_{AB}$~27 and 30 hr of integration time (Welikala et al. in prep). Upper panels (left-to-right): simulation input with six clumps at *z*~7; simulated *H*-band IFU cube for one spectral element (at *R*=4,000, between the OH lines) with spatial pixels of ~40 mas; same simulation with diffraction-limited spatial pixels, which over-resolves the clumps and lead to read-out limited spectra. Lower panel: *J*-band spectra (~40 mas spatial pixels) of the two brightest components, and composite of the four brightest; the black line represents the input spectrum and reveals two ISM lines in this wavelength range.

*IFU field-of-view :* The latest high-z, cosmological galaxy-formation simulations from Dayal et al. (2012) are known to reproduce the observed galaxy luminosity functions successfully, as well as the UV colours of galaxies at z~6-8 (Dunlop et al. 2013). An example result for the brightest galaxy in a simulated 10x10x10 Mpc$^3$ (co-moving) volume is shown in Fig. 5. The galaxy (based on many thousands of star particles in the high-resolution simulation) has a total absolute magnitude ($M_{UV}$) of −19.5 thus, at z~7.5, a predicted broad-band magnitude of $H_{AB}$ ~ 27.5. At the resolution of *HST* imaging, the object appears as one bright source at z~6.5 (left-hand panel of Fig. 5), but at z ~ 7.5 it appears as three isolated compact peaks (each with $H_{AB}$ ~ 28.5-29.5; central panel). The right-hand panel shows a smaller region in which additional structure can be seen at deeper magnitudes. More pertinently, the overlaid green box spans ~2x2 arcsec and contains >90% of the total *H*-band flux. Thus, these simulations argue that an IFU capable of collecting the light from a ~2 arcsec field would provide an essentially complete view of such a galaxy at z~7-7.5.

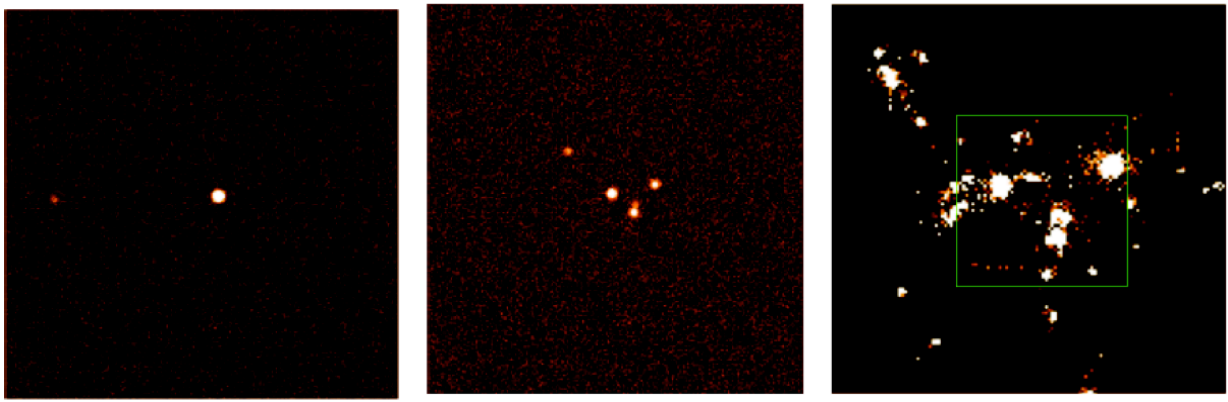

**Figure 5:** *Left-hand and central panels:* Simulated 12x12 arcsec *HST* WFC3/IR images of the brightest galaxy in a 10x10x10 Mpc (co-moving) volume simulation (from Dayal et al. 2012) at *z*~6.5 and 7.5, respectively. *Right-hand panel:* subset of the simulation at *z*~7.5 with ~0.04 arcsec pixels; the overlaid green square corresponds to 2x2 arcsec and encompasses most of the *H*-band flux.

# 3 SC2: SPATIALLY-RESOLVED SPECTROSCOPY OF HIGH-*z* GALAXIES

## 3.1 Following the assembly of galaxy mass as a function of look-back time

The availability of IFUs on 8-10 m class telescopes has heralded a new era of galaxy studies. We can now derive spatially-resolved kinematics and physical properties of distant galaxies at redshifts of up to *z*~3 (e.g., Förster-Schreiber et al. 2006, 2009; Yang et al. 2008; Law et al. 2009; Lemoine-Busserolle et al. 2010; Contini et al. 2012; Puech et al. 2012). However, above *z*~1.5-2 only the most massive, high-surface-brightness systems are within the grasp of current facilities. In addition, most of these samples have been assembled (necessarily) from a collection of selection criteria. Whether these samples are representative of the early Universe remains uncertain, but models of galaxy formation and evolution rely on analytic/empirical prescriptions, with inputs such as metallicity, angular momentum, and the spatial distribution of the gas taken from the existing observations (e.g., Hopkins et al. 2009; Dutton et al. 2012). This explains why the processes by which disk galaxies assembled their mass remain much debated and, in particular, the importance of minor vs. major mergers, as well as the accretion of hot vs. cold gas.

Making significant progress in our understanding of galaxy formation and evolution requires observations of substantial samples of galaxies, over a large enough volume to rule out field-to-field biases ('cosmic variance') and a large range of mass and redshift – only then will we be able to test the models rigorously with representative samples and reliable observables. The capabilities of the E-ELT will provide, for the first time, the potential for spatially-resolved observations of unbiased and unprecedented samples of high-*z* galaxies. Assembling a spectroscopic survey of hundreds of spatially-resolved high-*z* galaxies remains a strong scientific motivation for the E-ELT, and can only be obtained efficiently with a multi-IFU capability.

There are a wide range of opinions in the community regarding what processes drive galaxy evolution, and making progress in this area will require us to dissect a relatively large sample of intermediate-mass galaxies. Building a census of mass assembly and star formation in distant galaxies will require a combination of the deepest, high-resolution near-IR imaging (e.g. *HST*/WFC3, *JWST*) with an ELT-MOS. More specifically, with such a sample, it will be possible to:

- Infer the dynamical state of the observed galaxies, via the combination of spatially-resolved kinematics and deep imaging. Such a combination is essential to determine if a galaxy is a rotating disk (via comparisons wiith morpho-kinematic models of a regular rotating disk, e.g. Neichel et al. 2008), and to study internal galaxy structures (e.g., Forster-Schreiber et al. 2012; Law et al. 2012).

- Characterize the dynamical state of the surrounding halos (i.e., whether they are relaxed or not) by searching for companions within 100 kpc around each galaxy, via deep integrated-light spectroscopy (with a large multiplex). We will use relative velocities to model the trajectories of satellites, distinguish unbound satellites, and unveil major-merger pairs via their magnitude differences.

- Constrain the properties and dynamics of the stellar populations and ISM of the galaxies, including their companions, by comparing deep spectroscopy and spectral energy distributions (derived from multi-band imaging) with stellar population models. Further constraints on the source of ionization in the ISM gas will be derived using emission-line widths from 3D data (see Lehnert et al. 2009). The relative kinematics of the stellar and gaseous phase will be compared through potential shifts between absorption and emission lines, i.e., in/outflows will be detected and quantified (e.g., Rodrigues et al. 2012). [OII] line ratios will be used to derive electron densities (see Puech et al. 2006b), which will also constrain gas concentrations due to star formation or in/outflows.

- Derive metallicity gradients (using the $R_{23}$ index), which can be used as a tracer of gas accretion or mergers in distant galaxies (Cresci et al. 2010; Queyrel et al. 2012) as done in the Local Universe (Lagos et al. 2012).

- Characterize the evolution of the fundamental scaling relations. To first order, galaxies are self-gravitating systems, which can be described by at least three fundamental parameters, i.e., their (total) mass, velocity (including both bulk and random motions), and size. Scaling relations between these parameters or their proxies are therefore fundamental tools to characterize galaxy evolution. We will derive the integrated properties of all galaxies, including:
    ◦ their stellar mass, from both rest-frame colours and *K*-band luminosities (Bell et al. 2003) and SED fitting (e.g., Papovich et al. 2001];

- star-formation rate (combining UV and IR luminosities, e.g. Hammer et al, 2005);
- gas fraction, by inverting the Schmidt-Kennicutt law between star formation and gas densities; (e.g., Law et al. 2009; Puech et al. 2010);
- gas metallicity, using deep integrated spectroscopy and the $R_{23}$ estimator.

These will be used to study the evolution of scaling relations between mass and velocity (the stellar mass and baryonic Tully-Fisher Relations, Puech et al. 2010; Cresci et al. 2009; Vergani et al. 2012), mass and metallicity (the stellar mass-metallicity relation, e.g., Rodrigues et al. 2008, Manucci et al. 2009, as well as exploring the baryonic mass-metallicity relation), and between specific angular momentum and stellar/baryonic mass (Puech et al. 2007; Bouche et al. 2007).

- Use all these constraints to model all the galaxies in the sample, using hydro-dynamical models (e.g., Hammer et al. 2009; Hopkins et al. 2013). This step is crucial to disentangle what is driving star formation in each object and guiding us in discriminating between the possible disk origins. Such models can be used to derive the evolution of the merger rate as a function of redshift within a factor 2-3 in accuracy compared to theoretical predictions (Puech et al. 2012). Statistics will allow us to assess different possible mechanisms and give information on the cause of the star formation and mass build-up. The outcome of such a survey, such as the average gas fraction or the merger rate (e.g., Rodrigues et al. 2012), are invaluable constraints for semi-empirical models of galaxy evolution (e.g., Hopkins et al. 2009, 2010).

## 3.2    Requirements

*Observing modes and wavelength coverage:* The requirements for this case have been extensively discussed by Puech et al. (2008, 2010a). In brief, the goal is to target the rest-frame optical emission lines redshifted into the near-IR (see Fig. 6). An effective redshift limit to such studies is given by the [OII] emission line leaving the *K*-band (at *z*~5.6). However, *K*-band observations will require significantly larger exposure times than at shorter wavelengths because of the increased thermal background (Puech et al. 2010a). Without coverage of the *K*-band, observations would be limited to *z*~4 (*H*-band).

In summary, the essential requirements are the *YJ*- and *H*-bands (not simultaneously, but one band per observation), with the *K*-band as optimal/desirable. Spatially-resolving properties such as kinematics, metallicity, or SFR calls for an IFU capability. In addition, the case also calls for secondary, but highly desirable, single-object, large-multiplex deep near-IR spectroscopy to obtain information about satellite galaxies (redshifts and integrated properties), as detailed above.

*Multiplex:* Preliminary estimates of the number of targets of interest in the ~7 arcmin diameter patrol field were estimated to be tens to around a hundred galaxies, taking into account the spectroscopic success rate of measuring several emission lines between OH sky lines (Evans et al. 2010). Using results of the Design Reference Mission of the E-ELT, a survey of ~160 galaxies, spanning 2 ≤ z ≤ 5.6 in three redshift and mass bins, would require ~90 nights of observations (including overheads) with a multiplex of 20 IFUs (see Puech et al. 2010a). A similar survey limited to z≤4 (i.e. omitting the K-band) would provide observations of ~240 galaxies in ~12.5 nights. In contrast, such a sample with a single IFU instrument would require ~250 nights, needing 4-5 years of observations (assuming standard operations of sharing between instruments and programmes). Regarding the source density of satellite/dwarf galaxies suitable for integrated-light observations, a number of ~800 (1750) potential targets with $J_{AB}$ ≤ 26 (≤ 27) is found within the E-ELT patrol field (see details in SC3).

*Spatial resolution/image quality:* Most *z*>1 sources reveal clumpy morphologies (e.g., Elmegreen et al. 2005, 2007), which were proposed to be important steps of bulge formation (e.g., Bournaud et al. 2008). Such clumps are typically 1 kpc large, i.e., ~140 mas at *z*=4. Resolving these structures would therefore require at least ~70 mas/spaxel. However, recent results suggest that the clump lifetime is only 100-200 Myr (see Fig. 7; e.g., Forster-Schreiber et al. 2011; Wuyts et al. 2012). There are also theoretical reasons to suspect that feedback effects are indeed too strong for the clumps to survive long enough to sink to the centre and be a significant channel for bulge formation (Genel et al. 2011; Hopkins et al. 2012). In any case, on-going AO-assisted programmes on the VLT are currently assembling significant *z*~2 samples that will shed light on these issues (Mancini et al. 2011; Newman et al. 2012a,b), not to mention that the ELT-IFU that will be particularly well-suited to resolve these clumps at the diffraction limit of the E-ELT. We therefore argue that instead of optimising the spatial sampling of an ELT-MOS to resolve clump instabilities, a better compromise is to balance spatial sampling with surface brightness detection at the global galaxy scale.

Indeed, the next relevant spatial scale for studying distant galaxy structures is the diameter of the target galaxies, which is the scale at which large-scale motions are imprinted and which traces the dynamical nature of galaxies (e.g., Puech et al. 2008, see Fig. 8). Given the size of high-$z$ galaxies, this argues for spatial sampling in the range 50-75 mas (e.g. Puech et al. 2010a and references therein). More specifically, a ~75 mas pixel scale would allow to resolve sub-M* galaxies at $z$~4, or equivalently M* galaxies at $z$~5.6 (Puech et al. 2010a), while larger pixel scales would be too limiting on our ability to probe the mass functions at the largest redshifts (see Tab. 1). Simulations show that the required ensquared energy in 2x2 spatial pixels is ~30% (Puech et al. 2008; see Fig. 8).

*Spectral resolving power:* At least $R$=4,000-5,000 is needed to resolve the brightest OH sky lines and identify emission lines between them; this should also allow resolution of the [OII] doublet in the most distant galaxies.

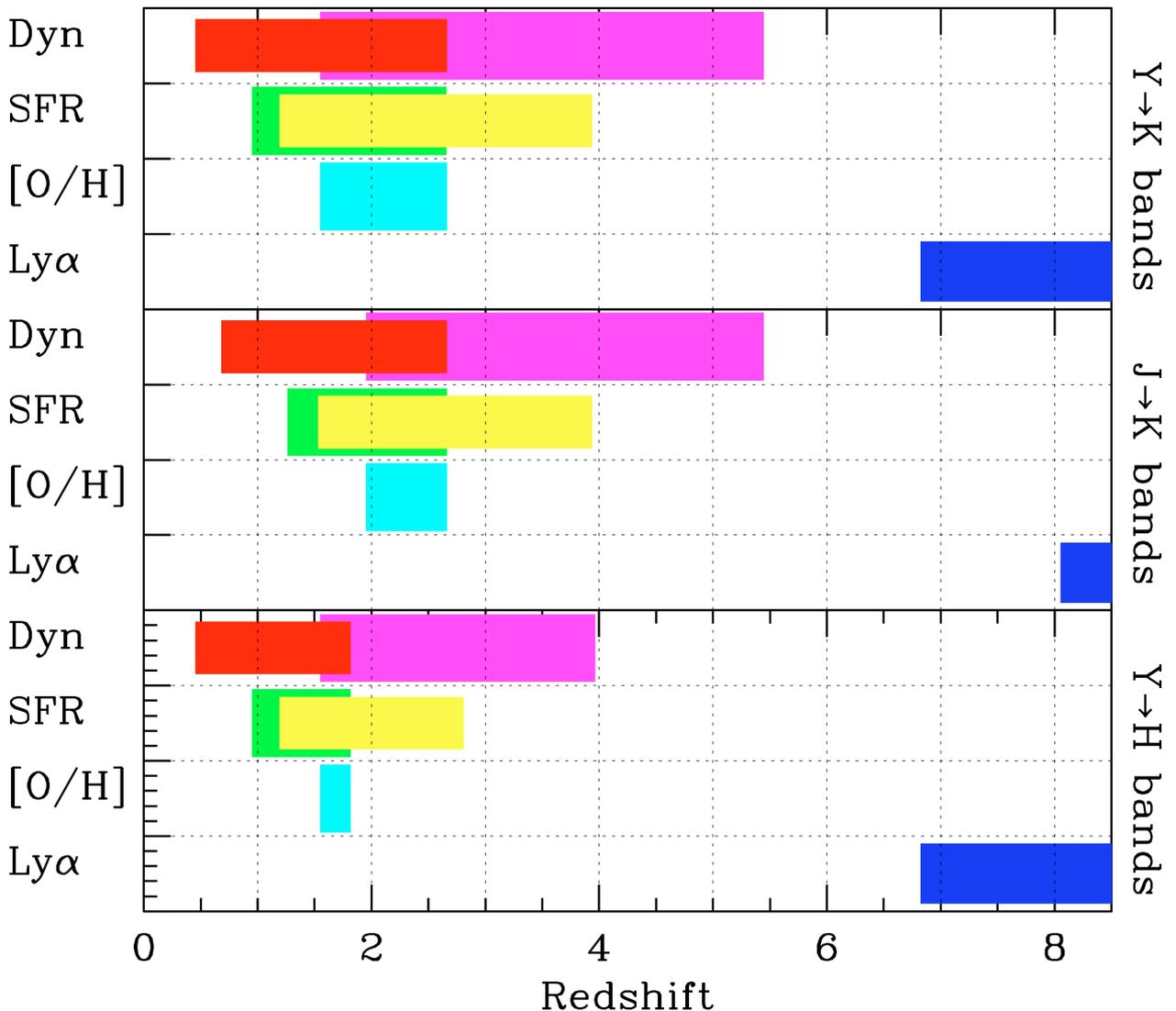

**Figure 6:** Physical quantities than can be determined from emission lines as a function of redshift and wavelength coverage. The following rest-frame lines were considered: H$\alpha$ $\lambda6563$, and [OII] $\lambda\lambda3726,3729$, for the dynamics (red and pink areas, respectively), H$\alpha$ and H$\beta$ $\lambda4861$ or H$\beta$ and H$\gamma$ $\lambda4341$ for the dust-corrected SFR (green and yellow areas, respectively), H$\alpha$, H$\beta$, [OII], [OIII] $\lambda5007$, and [OIII] $\lambda4959$ for the dust-corrected $R_{23}$-metallicity (cyan area). For comparison, the blue area shows at which redshifts one can target the Ly-$\alpha$ line (see SC1). The three panels show, from top-to-bottom, at which redshifts these physical quantities can be obtained as a function of the wavelength coverage (*Y* to *K* bands, *J* to *K* bands, and *Y* to *H* bands, respectively).

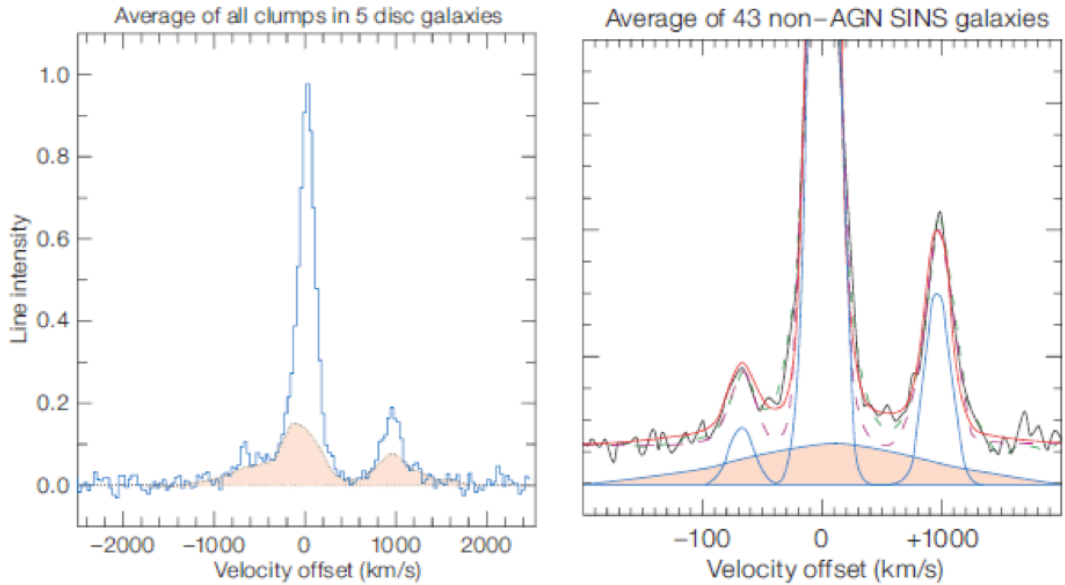

**Figure 7:** Detection of broad spectral features below emission lines in *z*~2 clumps (Förster-Schreiber et al. 2011) due to outflow winds. The inferred outflow rate is larger or equivalent to the clump SFR which, in turn, implies a ~100 Myr lifetime for the clumps.

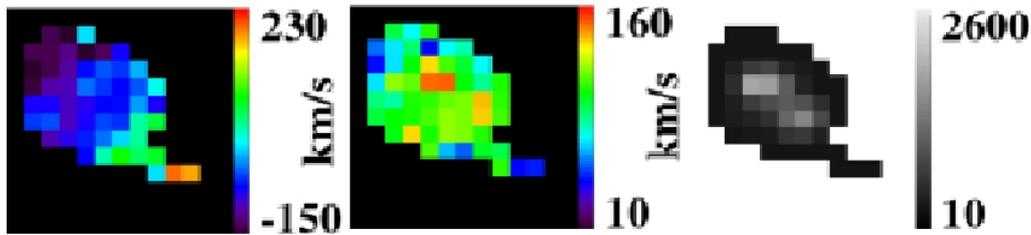

**Figure 8:** Simulated kinematics of a major merger at *z*~4 observed with an IFU of 75 mas/pixel (from Puech etal. 2008). From left-to-right: velocity field, velocity dispersion map, and [OII] emission line map, from which the kinematics was derived. The integration time was 24 hr and the ensquared energy within 150x150 mas² was 34%. The object total extension is 0.8 arcsec in diameter (~5.5 kpc). One can clearly see the two peaks corresponding to the two progenitors in the emission line map and velocity dispersion map.

**Table 1:** Evolution of the stellar-phase size as a function of redshift (*z*) and mass (expressed in fractions of M* at a given redshift). Sizes are expected half-light radii in the *K*-band in mas (see Puech et al. 2010a and references therein). Boxes with red/green characters are those for which 75 mas spatial sampling does not/does provide the (at least) two pixels per half-light radius ncessary to resolve the galaxies of given mass and redshift, respectively.

| z   | 0.1M*(z) | 0.5M*(z) | M*(z) | 5M*(z) | 10M*(z) |
|-----|----------|----------|-------|--------|---------|
| 2   | 170      | 300      | 380   | 670    | 850     |
| 4   | 80       | 150      | 190   | 330    | 430     |
| 5.6 | 70       | 130      | 160   | 280    | 350     |

*Spatial extent of each IFU:* Defined by the size of the galaxies on the sky and the need for good sky subtraction (e.g., A-B-B-A dithers within the IFU), this entails an IFU size of order 2"x2". Only the closest and most massive/luminous galaxies would require specific offset sky measurements but the relatively large collecting power of the E-ELT should limit the time penalty to reasonable total observing times of no more than a few hours depending on mass, size, and redshift.

Given, the very small size of dwarf galaxies, the best aperture for the single-object mode is that which maximises the signal-to-noise ratio for point-like sources in GLAO conditions. This was shown to be ~0.9" given the expected GLAO performance of the E-ELT (Navarro et al. 2010).

# 4 SC3: ROLE OF HIGH-*Z* DWARF GALAXIES IN GALAXY EVOLUTION

The deepest *HST* images have revealed a plethora of intrinsically faint galaxies at 1≤*z*≤3 (Ryan et al. 2007), providing a fairly steep low-mass slope of the UV luminosity function (Reddy & Steidel, 2009). This implies a considerable increase of the dwarf number fraction with redshift. Sub-L* galaxies may contribute to a significant fraction of the evolution in star-formation density, but we know little about their populations at high-*z*. Their faint apparent magnitudes ($m_{AB}$ ~ 24-27) prevent spectroscopy with current facilities and their morphological appearance is severely affected by cosmological dimming. Beyond understanding the overall role of sub-L* galaxies in the bigger picture of the evolution of mass and star-formation in the Universe, there are more specific questions which can only be addressed with an ELT-MOS, as illustrated by the following three examples.

## 4.1 The impact of low-surface-brightness galaxies (LSBGs) at high-z

LSBGs[1] represent 9% of the baryonic mass and a third of the HI mass density in the local Universe (Minchin et al. 2004), but their effect on the distant Universe is a complete mystery. They are gas rich and have average stellar masses comparable to dwarfs, such that they may contribute significantly to the increase of the number density of dwarfs at high-*z*. The gas fraction appears to increase rapidly with *z*, at a rate of 4% per Gyr from *z* = 0 to 2 (Rodrigues et al. 2012) and LSBGs might include a significant (dominant?) fraction of the Universal baryonic content at high-*z*. With high-multiplex observations we would investigate their ISM properties (dust, metal abundances, star-formation rates), while complementary IFU observations may provide their ionised gas fractions using inversion of the Kennicutt-Schmidt law.

## 4.2 HII galaxies to probe the curvature of the Universe (Λ)

With careful target selection[2], HII galaxies show a remarkably tight correlation between the luminosity of recombination lines and the velocity dispersion of the ionised gas, e.g., L(Hβ) vs. σ(Hβ), over four dex in luminosity (see Melnick et al. 1988, Bordalo & Telles, 2011). This arises when a starburst fully dominates its host galaxy by the correlation between ionising photons and the turbulence of the ionised gas; Melnick, Terlevich & Terlevich (2000) demonstrated it's validity up to *z*~3. Employing this correlation, Chavez et al. (2012) have used HII galaxies to provide an independent estimate of $H_0$ in the local Universe. Indeed, due to their brightness, HII galaxies at large redshifts (e.g., *z*~1.5-2.4) could provide larger observational samples than those of distant type Ia supernovae, at epochs for which the curvature effect is most pronounced. This could provide a complementary method to evaluate the dark energy equation-of-state at larger redshifts than currently probed by type Ia supernovae (Plionis et al. 2011). Deep exposures will be required to reach S/N~5 in the continuum of HII galaxies with $J_{AB}$ ~ 27 (requiring total exposure times of up to ~40 hrs).

## 4.3 The origin of dwarf galaxies

The signature of primordial dwarfs can be characterised from the peculiar properties of IZw18-type galaxies, whose spectra show prominent [OIII] 5007Å and Hα equivalent widths (>>1000Å) that severely contaminate the continuum, impacting on the shape of their SEDs. Deep near-IR imaging may provide targets down to $J_{AB}$ ≥ 29 and detection of their prominent emission lines will require MOS observations with the E-ELT, enabling determination of the evolution of the number density of primordial galaxies from *z* = 0-1.5 (or up to *z* = 2.4). Moreover, it has been claimed that most present-day dwarfs could have a tidal origin (Okazaki & Taniguchi, 2000). With an ELT-MOS this hypothesis can be probed directly at *z* = 0.8-2.4, the era at which most of the mergers would have occurred (e.g., Puech et al. 2012). The 3D locations of the dwarfs could be obtained relative to massive galaxies observed in a major-merger process. Their locations are expected to follow the tidal tails that can be modelled at the in-situ merger (e.g. Hammer et al. 2009), but this can only be tested via ELT spectroscopy. It is not clear whether such a mechanism would explain the observed density of dwarfs, thought it would be fascinating to explore them in the distant, gas-rich Universe. Another motivation for such a study would be to assess the number density of satellite galaxies and compare it to the expected number of haloes from the ΛCDM theory, from $z = 2.4$ to the present day. With target magnitudes of $J_{AB} \sim 26$ (27), an exposure time of 17 hrs (assuming GLAO) would provide S/N~8 (3) per continuum pixel.

---

[1] Typical magnitudes of $\mu_0(B)$>22 mag arcsec$^{-2}$ or $m_0(R)$>20.7 mag arcsec$^{-2}$ (Zhong et al. 2008).
[2] A nebular component which dominates the continuum flux, and with Gaussian emission components with $W_0(H\beta)$≥50Å.

## 4.4   Requirements

*Source densities:* The number of high-$z$ dwarfs with $M_{stellar} \leq 3\times10^9$ $M_\odot$ (i.e., masses comparable to the LMC or lower) with $J_{AB} \leq 26$ ($\leq 27$) within the E-ELT patrol field is ~800 (1750), arguing for a large multiplex capability (see Fig. 9) to build-up the large observational samples required.

*Spatial resolution/image quality:* This case is primarily (although not solely) concerned with exploiting the light-collecting power of the E-ELT and does not require spatially-resolved information for most of the target galaxies. Thus, the GLAO provided by the telescope as its standard operating mode will suffice.

*Wavelength coverage:* 0.8-1.7µm, enabling observations of [OII] 3727Å through to Hα out to $z \sim 1.5$, and [OII] 3727Å through to [OIII] 5007Å out to $z \sim 2.4$, thus enabling investigation of the properties of the ISM in these galaxies, i.e. dust content, metallicity, abundances, star-formation rates etc.

*Spectral resolving power:* Each of the above cases have similar requirements on spectral resolution. For instance, for observations of the H II galaxies, to reach the continuum level (to avoid contamination by stellar light) and to separate Gaussian Hβ profiles from those that are asymmetric, multiple, or rotationally-contaminated argues for $R \geq 5,000$ (with $R \sim 10,000$ desirable).

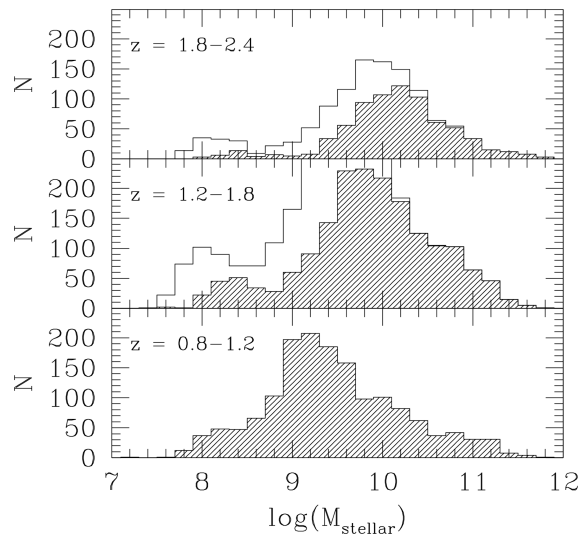

**Figure 9:** Distribution of stellar masses for galaxies selected by *HST*-WFC3 in the GOODS-S field within an area of 6.8'×10'. Photometric redshifts are derived from Dahlen et al. (2010; private comm.), absolute magnitudes are determined from an interpolation method (Hammer et al. 2001), and stellar masses are determined from the results of Bell et al. (2003). Shaded (and non-shaded in the upper panels) histograms show galaxies with $J_{AB} \leq 26$ ($\leq 27$) with 1,460 (3,100) dwarfs, with $M_{stellar} \leq 3\times10^9 M_\odot$.

# 5   SC4: Tomography of the IGM

The gas in the IGM is revealed by the numerous hydrogen absorption lines that are seen in the spectra of quasars bluewards of the Ly-α emission line from the quasar. It has been shown that the high-$z$ IGM contains most of the baryons in the Universe and is therefore the baryonic reservoir for galaxy formation. In turn, galaxies emit ionising photons and expel metals and energy through powerful winds which determine the physical state of the gas in the IGM. This interplay of galaxies and gas is central to the field of galaxy formation and happens on scales of the order of 1 Mpc (~2' on the sky at $z\sim2.5$, using standard cosmological parameters). The main goal of this case is to reconstruct the 3D density field of the IGM at

z~2.5 to study its topology, its chemical properties (e.g. HII-bearing sightlines), and to correlate the position of the galaxies with the density peaks.

The Ly-α forest seen in quasar spectra arises from moderate density fluctuations in a warm photo-ionised IGM. The spatial distribution of the IGM is related to the distribution of dark matter and the full density field can be reconstructed using a grid of sight-lines (Pichon et al. 2001; Caucci et al. 2008). A Bayesian inversion method can interpolate between the structures revealed by the absorption features in the spectra. Fig. 10 shows the level to which the matter distribution could be recovered, with 100 parallel sight-lines drawn through a 50×50×50 Mpc *N*-body simulation box for which synthetic spectra were then generated. From analysis of the simulated data, it was possible to recover structures over scales of the order of the mean separation of the sight-lines. To reach S/N ≥ 8 at $R$ = 24.8 (considering the background sources as point-sources), exposures of 8-10 hrs would be required per field, giving an ambitious programme of ~750 hrs to cover one square degree.

### 5.1.1 Requirements

*Source densities:* About 900 randomly distributed targets per square degree would be required to recover the matter distribution, with a spatial sampling of 0.5–2' at z~2. Once the density field is recovered, topological tests can be applied to recover the true characteristics of the density field. We thus expect observations of ~20 sources in a 7' field.

*Wavelength coverage:* This case pushes the blueward extent of the necessary wavelength coverage, requiring observations of the Ly-α forest. Thus, coverage down to at least 0.4 µm is essential, with coverage to 0.37 µm desirable. In addition, observing the objects redwards of the Ly-α emission would allow us to study the LBGs themselves, the metals in the IGM (in the case of quasars), and the quasars themselves if the IR is accessible.

*Spectral resolving power:* $R$ ≥ 5,000 is required, although $R$ = 10,000 would be helpful to avoid metal lines.

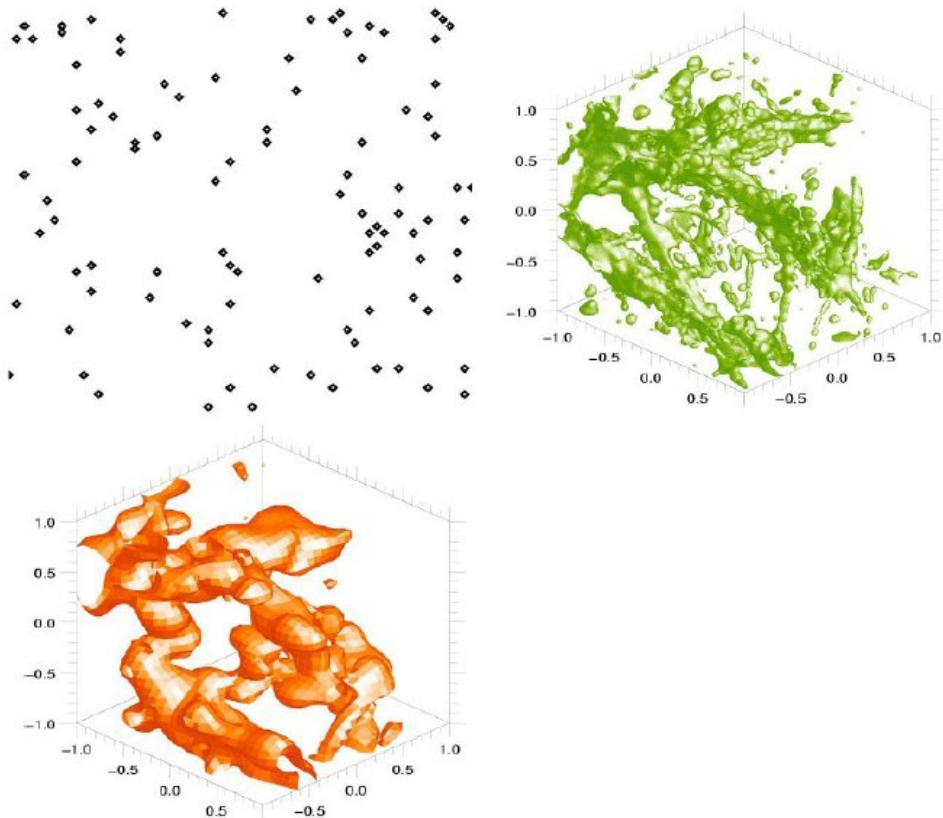

**Figure 10:** The input simulated density field is shown in the right-hand box (in green). One hundred (randomly-spaced) sight-lines to background sources are drawn through this box (upper-left). The corresponding spectra are used as the input data for the reconstruction, with the reconstructed density field shown in the lower box (in orange).

# 6 SC5: RESOLVED STELLAR POPULATIONS BEYOND THE LOCAL GROUP

Discoveries of disrupted satellite galaxies have demonstrated that our evolutionary picture of the Milky Way is far from complete (Majewski et al. 2003; Belokurov et al. 2006), let alone our understanding of galaxies elsewhere in the Universe. Deep imaging from ground-based telescopes and the *HST* has yielded colour-magnitude diagrams (CMDs) with unprecedented fidelity, providing new and exciting views of the outer regions of galaxies beyond the Milky Way (e.g. Ferguson et al. 2005; Barker et al. 2007; Richardson et al. 2008). From comparisons with stellar evolutionary models, these data enable us to explore the star-formation and chemical-enrichment histories of the targeted regions, providing a probe of the past evolution and, in particular, the merger/interaction histories of these external galaxies. New constraints on galaxy formation models have also been provided by evidence for larger-scale structure and alignment of the galaxies in the Milky Way (Pawlowski et al. 2012) and Andromeda groups (Ibata et al. 2013).

Although photometric methods are immensely powerful when applied to extragalactic stellar populations, only via precise chemical abundances and stellar kinematics can we break the age-metallicity degeneracy, while also disentangling the populations associated with different structures, i.e. follow-up spectroscopy is required. Over the past decade the Calcium II Triplet (CaT, spanning 0.85-0.87 µm) has become an increasingly used diagnostic of stellar metallicities and kinematics in nearby galaxies (e.g. Tolstoy et al. 2004). However, 8-10m class telescopes are already at their limits in pursuit of spectra of the evolved populations in external galaxies, e.g. Keck-DEIMOS observations in M31 struggled to yield useful S/N below the tip of the red giant branch (RGB) at $I > 21.5$ (Chapman et al. 2006). To move beyond the Local Group – to investigate whether similar processes are at work in other large galaxies, and what role environment and galaxy morphology have on galaxy evolution – we need the sensitivity of the E-ELT.

With its vast primary aperture and excellent angular resolution, the E-ELT will be *the* facility to unlock spectroscopy of evolved stellar populations in the broad range of galaxies in the Local Volume, from the edge of the Local Group, out to Mpc distances. This will bring a wealth of new and exciting target galaxies within our grasp, spanning a broader range of galaxy morphologies, star-formation histories and metallicities than those available to us at present. These observations can then be used to confront theoretical models to provide a unique view of galaxy assembly and evolution. There are many compelling and ground-breaking targets for stellar spectroscopy of individual resolved stars with the E-ELT including:

- NGC 3109 and Sextans A with sub-SMC metallicities (< 0.2 solar), both at 1.3 Mpc;
- The spiral dominated Sculptor 'Group' at 2-4 Mpc
- The M83/NGC5128 (Centaurus A) grouping at ~4-5 Mpc;

Beyond these targets, the E-ELT will also important for spectroscopy of the (less resolved) stellar populations in more distant galaxies, such as NGC 3379 (at ~11 Mpc) and systems in the Virgo Cluster.

## 6.1 Spectroscopy of evolved stars in the Sculptor Galaxies

The top priority case in this area is a 'Large Programme'-like survey of the evolved populations in the Sculptor Group, comprising five spiral galaxies (NGC 55, NGC 247, NGC 253, NGC 300 and NGC 7793) and numerous dwarf irregulars. Distance estimates over the past decade have revealed that this 'group' is actually two distinct components (e.g. Karachentsev et al. 2003), at approximately 1.9 Mpc (NGC 55 & 300) and 3.6-3.9 Mpc (NGC 247, 253 & 7793). These five galaxies represent the most immediate opportunity to study the star-formation history and mass assembly of spirals beyond the limited sample available at present, i.e. the Milky Way, M31 and M33. Their masses are in the range $1.5\text{-}8 \times 10^{10}$ M$_\odot$, putting them on a par with M33 – it is exactly these late-type, low-mass, small bulge (or even bulge-less) spirals that theoretical N-body/semi-analytic simulations struggle the most to reproduce (D'Onghia & Burkert, 2004).

Two of the Sculptor galaxies are shown in Fig. 11 – these are large extended galaxies in which there is a wide range of stellar densities/crowding. A key point to note is that there is already substantial deep imaging available of such galaxies from, e.g., the *HST*; i.e. we already have catalogues of potential targets, but lack the facilities to obtain spectra with adequate S/N. A HARMONI-like instrument will be well suited to spectroscopy of stars in individual dense regions in external galaxies (and the Milky Way), but the larger samples needed to explore *entire* galaxy populations will require an ELT-MOS. To recover the star-formation histories and structures in the Sculptor galaxies we advocate a two-fold approach (depending on the target source densities), as discussed in the next two sections.

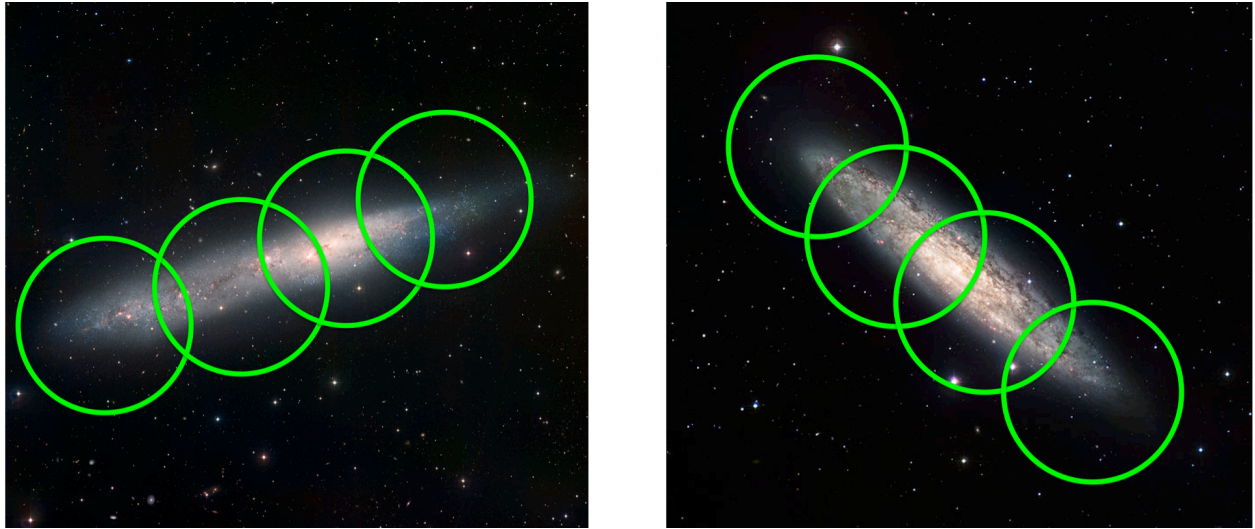

**Figure 11:** Illustrative 7' diameter ELT-MOS pointings in NGC 55 at 1.9 Mpc (*left*) and NGC 253 at 3.6 Mpc (*right*).

## 6.2  Requirements: Halos of external galaxies

*Spatial resolution/image quality:* The outer regions of external galaxies can provide some of the most dramatic evidence about their past star-formation histories, via the presence of extended structures, stellar streams etc. The objective is to observe hundreds of stars per galaxy but, given the low source densities at large galactocentric distances, high-performance AO is not required and the GLAO correction enabled by M4 and M5 will provide acceptable image quality.

*Wavelength coverage:* We assume the use of the CaT as a metallicity diagnostic, as used by the *Gaia* mission and as used in many studies in the Galaxy and other galaxies in the Local Group. At large galactocentric distances we would expect relatively metal-poor populations, but an important characteristic of the CaT is that its three absorption lines are relatively strong, meaning that its relationship to metallicity, [Fe/H], is robust over a large range (e.g. Cole et al. 2004; Carrera et al. 2007), with an updated calibration from Starkenburg et al. (2010) valid to metallicities as low as [Fe/H] ~ −4. Other strong spectral features such as the Mg I b triplet or the G-band may also be used as additional diagnostics.

*Spectral resolving power:* Battaglia et al. (2008) demonstrated that, with careful calibration and S/N ≥ 20 Å$^{-1}$, metallicities obtained from the low-resolution mode of FLAMES-Giraffe (*R* ~ 6,500) are in agreement with direct measurements from the high-resolution mode (*R* ~ 20,000). In practice, *R* ≥ 5,000 is sufficient, given adequate S/N (i.e. ≥ 20).

## 6.3  Requirements: Disc regions of external galaxies

*Wavelength coverage:* While, the CaT is a widely-used spectral feature for studies of stellar populations, the AO correction will be more effective at longer wavelengths, and it is worth considering other diagnostics. Observations in the *J*-band (at *R* of a few thousand) were demonstrated by B. Davies et al. (2010) to provide robust estimates of metallicities of red supergiants (RSGs). The 1.15-1.22 µm region includes absorption lines from Mg, Si, Ti, and Fe so, while the lines are not as strong as those in the CaT, they provide direct estimates of metallicity. The potential of this region for extragalactic observations of both RGB stars and RSGs with the E-ELT was further investigated by Evans et al. (2011). They concluded that a continuum S/N > 50 (per two-pixel resolution element) was required to recover simulated input metallicities to within 0.1 dex, sufficient for many extragalactic applications. On-sky tests of the *J*-band methods are underway via VLT–XShooter observations of RSGs in the Magellanic Clouds. Once the additional problems of real observations are taken into account (e.g. telluric subtraction), analysis of these data suggests that the required S/N for good metallicity estimates is more like 100 (Davies et al. in press).

*Spatial resolution/image quality:* This becomes a significant factor in the denser regions of external galaxies, and therefore makes stronger demands on the AO performance. For instance, the spatial resolution delivered by *HST*-ACS and WFC3 observations is more than a factor of three finer than the sampling provided by the GLAO correction – to successfully follow-up such observations (for example), will require additional correction. The WEBSIM tool developed by Puech et al. (2008) was used to simulate performances for *J*-band stellar spectroscopy with MOAO PSFs (see Evans et al. 2011). The results in the left-hand panel of Fig. 12 show the mean S/N (and standard deviation) from ten simulations for individual stars with $J$ = 22.75 and 23.75. The results in the right-hand panel of Fig. 12 shown the S/N recovered from two configurations of natural guide stars (NGS), which were representative of the two likely extremes of MOAO performance from different asterisms.

In sheer survey speed and sensitivity for one pointing, the maximum efficiency for such point-source observations would be achieved with spatial sampling of 20 mas; marginally longer exposures would be required for spatial sampling of 30-40 mas. The maximum spatial sampling acceptable is ~75 mas – equivalent to critically-sampled optical images from *HST,* and short-wavelength near-IR images from *JWST*-NIRCAM. With 75 mas spatial pixels the exposure time to reach the same S/N as the 37.5 mas pixels adopted in the EAGLE study would need to be 2-2.5 times longer. However, if the same number of detector pixels were used by each IFU, coarser sampling would enable IFUs with an area twice as large (leading to observations of more stars per IFU). Thus, in terms of overall survey speed *for a given number of stars* across the full extent of an external galaxy, either would be acceptable (and in fields where the spatial resolution is more critical, individual pointings with a HARMONI-like IFU could be obtained).

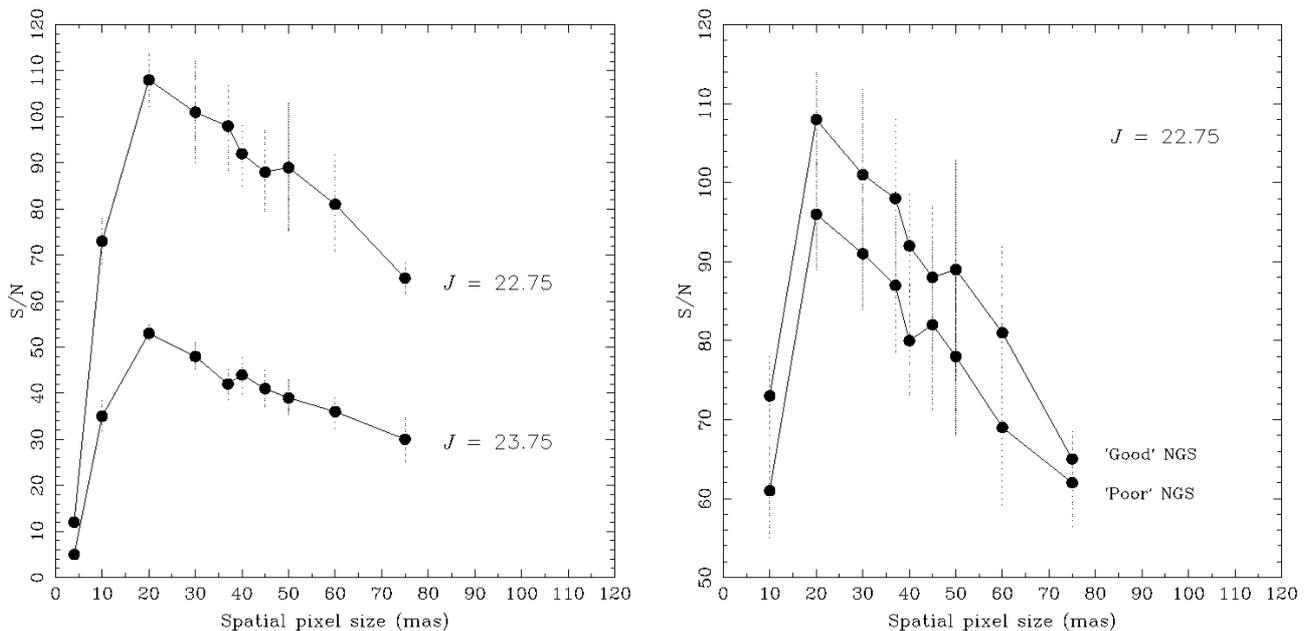

**Figure 12:** Simulated performances (S/N per 2-pixel resolution element) for individual stars (with MOAO correction). *Left-hand panel:* Illustrative performances for targets with $J$ = 22.75 and 23.75 for MOAO with a 'good' configuration of natural guide stars (NGS). *Right-hand panel:* Comparison of performances for $J$ = 22.75 with 'Good' and 'Poor' NGS configurations (note the different range plotted in S/N cf. the left-hand panel).

*Multiplex:* While the objective is to compile spectroscopy of individual stars, the extended spatial coverage provided by IFUs is an attractive means to obtain large samples. IFUs with fields on the sky of ≥1.0"x1.0" would provide adequate spatial pixels for good background subtraction combined with multiple stars per IFU. Indeed, with potentially tens of stars per IFU, the effective multiplex can be large. We require samples of >>100 stars per galaxy, to adequately sample their different populations and sub-structures (thus requiring multiple E-ELT pointings across each galaxy, e.g. Fig. 11). To assemble ~1,000 stars per target galaxy (within a few pointings) suggests a multiplex in the range of 10 to 20 IFUs. Of course, individual fields in selected galaxies could be observed with a HARMONI-like IFU, but to sample the full range of structures would entail a Large Programme per galaxy (which is unrealistic given the other demands on such an instrument and the E-ELT overall). An ELT-MOS could address the science goals in a few pointings and within a Large Programme could investigate all of the major Sculptor galaxies (not to mention observations in galaxies further away, e.g. Cen A, M83, etc).

## 6.4    Comparison of *J*-band and CaT performances

Evans et al. (2011) presented WEBSIM simulations of MOAO observations of cool stars using both the CaT and the *J*-band diagnostics. We have also simulated new CaT observations with coarser spatial sampling from GLAO using a version of WEBSIM for simulations of EVE performances, adopting similar parameters as Evans et al. (2011) to enable comparisons: a total throughput of the telescope of 80%, an instrumental throughput (including the detectors) of 35%, a low read-out noise of 2e⁻/pixel, and an exposure time of 20×1800s. Other inputs were a spatial sampling of 0.3" (of a 0.9" aperture), $R = 5,000$, and one of the *I*-band GLAO point-spread functions (PSFs) from Neichel et al. (2008). The PSF included a ring of nine NGS at a diameter of 7' – the configuration of guide stars (both natural and laser) will be somewhat different to this, but this PSF serves as a first-order test of the likely performances in the GLAO case (excluding the known limitations in the simulations).

Taking the results from Table 4 of Evans et al. (2011) a S/N ~ 100 is achieved for *J* ~ 22.5 depending on the AO asterism and observing conditions/zenith distance; this sensitivity estimate also takes into account the known limitations in the simulated MOAO performances. From ten GLAO simulation runs to calculate CaT spectra we find a continuum S/N = 21 ± 3 (per pixel) for *I* = 23.5, sufficient to recover the metallicity (cf. Battaglia et al. 2008). The intrinsic *I* − *J* colours for RGB stars are ~0.75 mag so, to first order, the GLAO observations of the CaT and the MOAO observations in the *J*-band provide sufficient S/N to recover the metallicity of a given RGB star – i.e. they are roughly competitive for a given target (ignoring the effects of extinction and crowding for now).

MOS spectroscopy on the E-ELT will provide the capability to determine stellar metallicities out to Mpc distances (for stars near the tip of the RGB, as shown in Fig. 13) and, in the case of the AO-corrected *J*-band observations, out to tens of Mpc for RSGs. The latter case will open-up a huge number of external galaxies, spanning all morphological types, for direct study of stellar abundances and calibration of the mass-metallicity relation, as shown in Fig. 14.

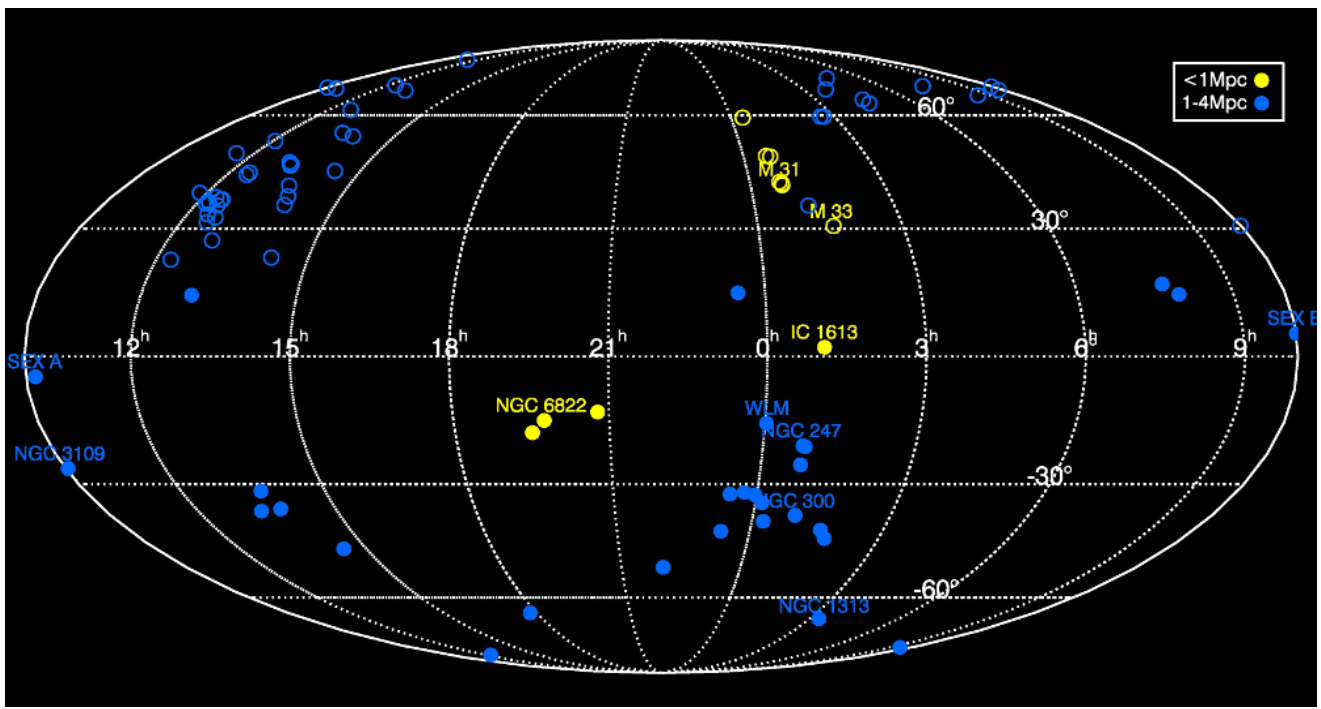

**Figure 13:** MOS spectroscopy of evolved red giants with the ELT will be possible out to distances of several Mpc, opening-up a wide range of external galaxies for direct abundance determinations. Galaxies with δ ≤ 20°, i.e. those observable from Cerro Armazones at reasonable altitudes (≥45°), are indicated by the closed symbols.

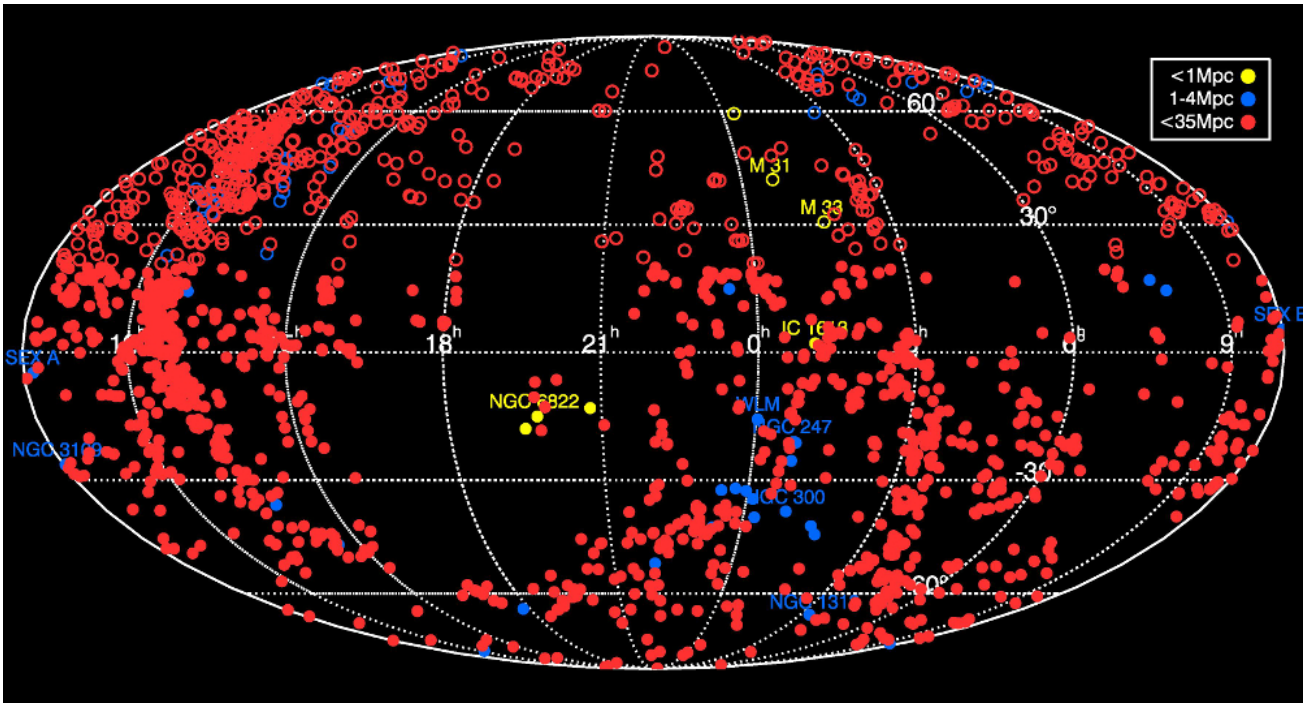

**Figure 14:** MOS spectroscopy of RSGs with the ELT will open-up an even larger range of external galaxies for direct abundance determinations. Galaxies with δ ≤ 20°, i.e. those observable from Cerro Armazones at reasonable altitudes (≥45°), are indicated by the closed symbols.

We propose that GLAO observations of the CaT will be sufficient for observations in the sparse regions of external galaxies such as those in Sculptor, and that MOAO observations in the *J*-band are well suited for investigation of the main bodies of the target galaxies. Provision of both of these modes is highly complementary, probing different regions and populations (thus providing good sampling of each spatial and kinematic feature). This combination of modes will give a complete view of each galaxy, which is required to confront models of star-formation histories and past interactions. This provides a relatively 'clean' split in the required wavelength coverage for the two modes at ~1µm although, pending more detailed results of the *J*-band methods, extension of the wavelength coverage of the IFUs to 0.8µm is retained as a goal.

# 7 SC6: GALAXY ARCHAEOLOGY WITH METAL-POOR STARS

The Universe that emerged from the hot and dense phase after the Big Bang had an extremely simple chemical composition: stable isotopes of hydrogen and helium, with traces of $^{7}$Li. As discussed in SC1, the source of the photons which reionised the early Universe is still unclear. A leading contender is ionising UV photons from the first stars which formed from the primordial gas (so-called 'Population III' stars). A consequence of the extremely low metallicity is that this first generation of stars is expected to be comprised of very massive stars (e.g. Bromm & Larson, 2004) – metallic species are thought to play an essential role in cooling while the natal gas undergoes collapse and, without these vital cooling agents, fragmentation of the gas is probably inhibited, leading to a unique and relatively massive first generation of stars. The nuclear reactions at work in the stellar interiors of this first generation of stars will then start manufacturing metals, with the result that the pristine ISM will be 'chemically polluted' within a few stellar generations as the massive stars explode as supernovae.

This theoretical paradigm requires observational confirmation, e.g., searching for emission features from (rest-frame) He II 1640Å as mentioned in SC1. In particular, we want to know the form of the initial mass function (IMF) for this first generation of stars, i.e. what mass range did it span? Where these first stars all very massive (a 'top-heavy' IMF), or were there also lower-mass stars formed from the primordial gas? Moreover, recent numerical simulations suggest that the distribution of possible stellar masses in this epoch may be much broader than previously thought, extending down to ≤ 1$M_{\odot}$ (e.g., Clark et al. 2011; Greif et al.

2011). If stars with M ≤ ~ 0.8M$_\odot$ were actually formed a few hundred Myr after the Big Bang, they should still be shining today as their lifetime is larger than the Hubble time.

We can constrain the issues of star-formation in very metal-poor environments by analysing the metallicity distribution functions (MDFs) of stellar populations in the local Universe. One of the best tracers to determine these MDFs are stars at the main-sequence turn-off (MSTO), which are both numerous and relatively uncontaminated in terms of other populations, are easy to select on the basis of their colours, and are the brightest population among the long-lived, chemically-unmixed stars, i.e. their chemical abundances are relatively unaltered since their formation. To advance this field we need the sensitivity of the E-ELT, combined with both a large multiplex (to compile large samples of stars) and sufficient spectral resolving power for detailed chemical analysis ($R \geq 20,000$). The relative proximity of the targets means that the GLAO correction provided by the E-ELT is sufficient in terms of the required angular resolution.

## 7.1  Metal-poor stars: the state-of-the-art in 2013

### 7.1.1  The Galaxy

Significant effort has been invested over the past few decades in searching for primordial stars in the Galaxy. These stars are the long-lived descendants from the earliest stellar generations, and will have formed from a (near-)pristine ISM, which would have only been weakly enriched in metals from the first supernovae. Their atmospheres therefore give us a fossil record of the ISM from which they were formed, corresponding to redshifts of $z \geq 10$. Having a direct tracer of chemical abundances at such an early time can provide fundamental constraints on the properties of the first generations of star formation, as well as giving indirect information on their masses and ionising feedback (of interest in the context of reionisation).

Until recently, the most metal-poor stars known were four giants and one turn-off binary star with $-4.1 \leq$ [Fe/H] $\leq -3.7$ (Norris et al. 2000; François et al. 2003; Cayrel et al. 2004; Gonzalez Hernandez et al. 2008), and the deepest survey searching for metal-poor stars was the Hamburg-ESO survey (HES, Christlieb et al. 2008), going down to $V = 16$. The MDF derived from the HES has a vertical drop at [M/H] = −3.5 (Schőrck et al. 2009), in excellent agreement with predictions that a 'critical metallicity' of [M/H] = −3.5 is required for the formation of low-mass stars. Above this value, cooling from the fine-structure lines of ionised carbon and neutral oxygen is thought to be sufficient for the formation of low-mass stars (e.g. Bromm & Loeb, 2003; Frebel, Johnson & Bromm, 2007). The discovery of metal-poor stars with strong C and O enhancements provided support to this theory (Christlieb et al. 2002; Frebel et al. 2005; Norris et al. 2007), but the discovery that the extremely metal-poor star SDSS J102915+172927 is not strongly enhanced in C or O challenges this model (Caffau et al. 2011). Indeed, an alternative scenario predicts that dust cooling plays the dominant role and argues for a lower critical metallicity of [M/H] ~ −5 (e.g., Schneider et al. 2006; Dopcke et al. 2011).

Work is now underway to improve our understanding of the low-metallicity tail of the MDF – major progress can be expected in the coming years using data from the Sloan Digital Sky Survey (SDSS, York et al. 2000) which has obtained low-resolution spectra for >$10^5$ stars, providing an excellent opportunity to search for primordial stars. The low resolution of the SDSS spectra limits metallicity determinations to better than ~1 dex, but observations at higher resolution are now underway at the VLT to calibrate the large amount of SDSS data, to then confront the competing theories.

### 7.1.2  The Galactic bulge

The stellar populations of the Galactic bulge are a template for studies of ellipticals and bulges of spirals. Its formation can give hints on proto-galaxy counterparts observed at high-redshifts, making the detailed study of this component of our Galaxy of great importance. Its stellar populations show a metal-poor component of metallicity extending at least down to [Fe/H] = −1.1 (Hill et al. 2011), and perhaps even lower (González et al. 2013), which is probably the oldest component of the Galaxy. The kinematics of the metal-poor population is compatible with that of an old spheroid, while the more metal-rich population is compatible with a bar. This suggests distinct formation scenarios for the two (Hill et al. 2011).

Current observations are limited to giant stars, but in order to disentangle the complex mix of stellar populations in the Galactic Bulge, observations of dwarfs would give important information on the field population and particularly on stars in globular clusters. Typical MSTO magnitudes of stars in the bulge are: for Baade's Window $V_{MSTO}(BW) = 19.5$, for the metal-rich globular NGC 6528 $V_{MSTO}(NGC6528) = 20.8$ (Ortolani et al. 1995) and for the metal-poor globular NGC 6522 $V_{MSTO}(NGC6522) = 20.4$ (Piotto et al. 2002; Barbuy et al. 2009). A few clusters are brighter than these examples, but many others are fainter.

VLT-FLAMES spectroscopy at $R\sim20,000$ in the optical can reach $V=17$ in about 2 hours for a reasonable S/N. In the near-IR it is possible to reach 2 to 3 magnitudes deeper in the sense that the SEDs of these stars peak at longer wavelengths. However, the near-IR lines available ido not include some key elements: lines of FeI and FeII of varied excitation potential are only found in the optical, as well as the lithium line, and many lines of heavy elements. It is interesting to note that Bensby et al. (2010) observed lithium in a metal-poor Bulge dwarf (which was brightened by a micro-lensing event), finding it to be compatible with the abundance of the Spite plateau in the Galactic Halo. It is difficult to draw strong conclusions from the observations of a single star, yet the discovery of a Spite plateau (see next section) in the metal-poor Bulge population may have far-reaching cosmological implications. An ELT-MOS will be able to obtain optical spectroscopy of dwarf and turn-off stars in the Bulge, allowing us to compile statistically significant samples of stars to address specific problems such as the existence of the Spite plateau and/or the evolution of the heavy elements in this important Galactic component.

### 7.1.3 The galaxies of the Local Group

It is important to know if the star-formation process in other galaxies resembles that in the Galaxy (see review by Tolstoy, Hill & Tosi, 2009). Interestingly, the MDFs of the dwarf spheroidal galaxies obtained to date are significantly different from that in the Galaxy (see Fig. 15); the metal-poor tail of the Galactic MDF is significantly more populated than in the dwarf spheroidals, but it is not clear why this is the case.

The MDFs available for nearby external galaxies have been obtained from [Fe/H] measurements in giant stars, generally employing the CaT calibration (e.g., Battaglia et al. 2006, Helmi et al. 2006). The improved CaT calibration from Starkenburg et al. (2010) is valid over the range $-4 \leq$ [Fe/H] $\leq -0.5$ and indicated that the numbers of extremely metal-poor stars in these external galaxies were actually not as low as thought previously. This was confirmed by precise measurements from high-resolution VLT-UVES observations of five metal-poor candidates in three dwarf spheroids (Fornax, Sculptor, and Sextans), finding [Fe/H] $\leq -3$ in each case (Tafelmeyer et al. 2010).

Another interesting aspect of this field is related to the ground-breaking discovery by Spite & Spite (1982) that Galactic metal-poor stars at the MSTO have a constant Li abundance irrespective of their metallicity or effective temperature (the so-called 'Spite Plateau'). The most straightforward interpretation of this result is that the observed Li is primordial, i.e. produced during the first three minutes of the Universe. As mentioned earlier, in those earliest moments only nuclei of deuterium, two stable He isotopes ($^3$He and $^4$He), and $^7$Li were synthesised. Their abundances depend on the baryon/photon ratio, thus on the baryonic density of the Universe. In principle, the Spite Plateau allows us to determine the baryon/photon ratio, which can not be deduced from first principles.

This interpretation of the Spite Plateau is seriously challenged by the measurement of the baryonic density, with unprecedented precision, from the fluctuations of the cosmic-microwave background by *WMAP* (Spergel et al. 2007). However, we have indications that the plateau is Universal. For instance, Monaco et al. (2010) observed the Spite Plateau in ω Cen, which is generally considered to be the nucleus of a disrupted satellite galaxy. It is of paramount importance to verify if this is the case in other Local Group galaxies, which possess metal-poor populations, but spectroscopy of extragalactic stars at the MSTO requires the sensitivity of the E-ELT.

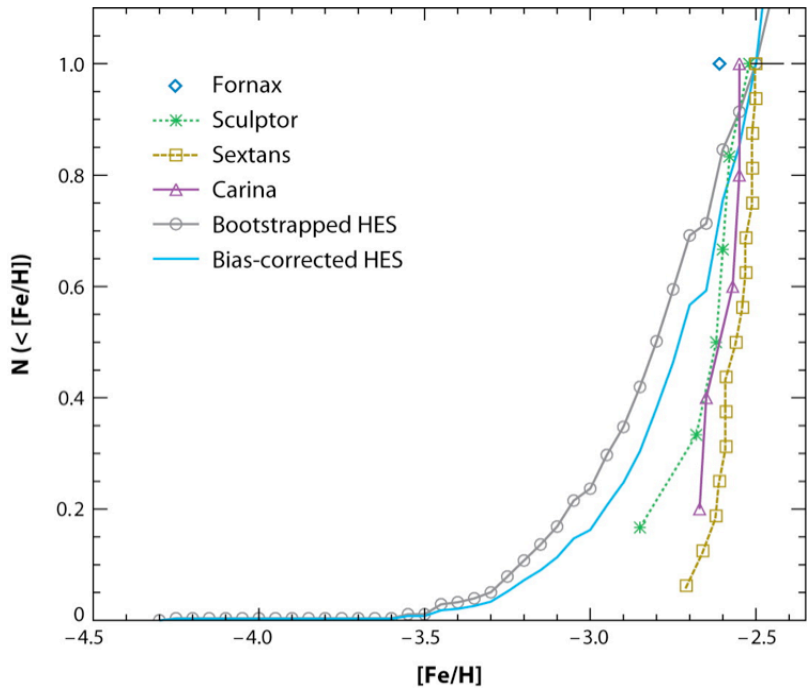

**Figure 15:** Galactic metallicity-distribution functions (MDFs) from the Hamburg-ESO Survey (HES; bias-corrected results are from Schőrck et al. 2009) compared to MDFs for four dwarf spheroidals from the DART survey (Helmi et al. 2006); this is Fig. 10 from Tolstoy, Hill & Tosi (2009). If one uses the recalibration of CaT metallicities from Starkenburg et al. (2010) this figure changes somewhat, but significant differences persist between results for the Galaxy and its dwarf satellites.

## 7.2 Goals for the E-ELT

One the key questions in the Local Group is whether the galaxies were formed from the early primordial gas, or if they formed from gas that had already been partially enriched, thus providing a metallicity 'floor'. This has important cosmological implications. In the standard hierarchical scenario the first structures to form are dwarf galaxies which subsequently merge to form larger structures like the Galaxy and other massive disc galaxies. If the MDFs of Local Group dwarfs display a clear metallicity 'floor', then either the hierarchical galaxy formation model is wrong, or the present-day 'surviving' dwarf galaxies formed later (in which scenario, they are not the relics of the primordial dwarfs).

The current limiting factor in extragalactic studies is that only giant stars are bright enough to have high-quality, high-resolution spectroscopy with an 8-m class telescope (e.g. giant-branch stars in 'nearby' dwarf spheroidals have $V \leq 18$). Unfortunately, the statistics available from the analysis of extragalactic RGB stars is not sufficient to determine the metal-poor tail of the MDF robustly in their host galaxies. If the fraction of stars below [Fe/H] ~ −4 is $10^{-4}$, then at least $10^4$ stars need to be observed to find just one. There are simply not enough giant-branch stars in most of the Local Group dwarf galaxies to sample these rare populations, and the MDFs determined to date are likely to be severely incomplete in their metal-poor tails. As mentioned earlier, the best stars for determining a well-sampled MDF are those at the MSTO, and to observe these faint stars in extragalactic systems we need the sensitivity of the E-ELT.

We require large samples of stars at the MSTO, in multiple galaxies (to further improve the statistics). The *V*-band magnitudes of stars at the MSTO in nearby galaxies are summarised in Table 2; thus we require high-resolution, absorption-line spectroscopy down to *V*~25. An additional crucial piece of information will be carbon abundances of these metal-poor populations – if they are carbon-enhanced this would indicate that the star-formation paradigm advocated by Bromm & Loeb (2004) is dominating the process, whereas (mainly) carbon-normal abundances would favour star formation driven by dust cooling and fragmentation. Carbon abundances can be determined for MSTO stars from observations (at $R \geq 10,000$) of the G-band feature (at 430 nm).

## 7.2.1 Instrument requirements

Observations are required of stars at the MSTO in Local Group galaxies down to $V\sim25$. Effective temperatures of each target will be derived from the photometry used to select candidates, as well as from fits to the wings of the Hα line at 656.3 nm (giving sufficient precision of $\sim \pm200$ K). Stellar gravities will also be derived from the photometry, but can be further constrained from spectroscopy.

*Spectral resolving power:* High-resolution spectroscopy is required at $R \geq 20{,}000$ (optimal), $R \geq 15{,}000$ (essential) to determine accurate chemical abundances/metallicities, but only selected wavelength regions are required.

*Wavelength coverage:* The optimum optical regions for high-resolution spectroscopy are well understood from the extensive work already completed with, e.g., VLT-FLAMES and VLT-UVES. Abundance estimates are required for: Fe, α-elements (e.g. Mg, Ca), C (from the G-band), N (from the A-X CN band), Li, Ba, Eu, and Sr. Thus, the optimal wavelength ranges required are 380-520 nm and 640-676 nm, with an essential requirement on the bluewards range of 410-460 nm. If these are obtained simultaneously then that would save by a factor or two in total observing time.

Table 2: *V*-band magnitudes of the main-sequence turn-off (MSTO) in selected nearby southern galaxies

| Galaxy | $V_{MSTO}$ (mag) |
|---|---|
| Sagittarius dwarf | 22.4 |
| LMC | 23.9 |
| SMC | 24.5 |
| Sculptor dwarf | 24.8 |
| Sextans dwarf | 24.8 |
| Carina dwarf | 25.0 |
| Fornax dwarf | 25.7 |

# 8 ADDITIONAL CASES

We have introduced six top-level cases, which will drive the design of an ELT MOS, but these are not the limits of its scientific potential. We now briefly introduce four topics which also require an ELT-MOS, by way of illustrating the huge scientific applications of such a facility. Of course, these are not the full scope of the potential MOS science on the E-ELT, with many other cases still to be investigated in more detail, e.g., distant galaxy clusters, other Galactic stellar populations (such as white dwarfs), extragalactic stellar clusters, etc.); i.e. an ELT-MOS will address a truly broad range of scientific goals!

## 8.1 Testing the invariance of the extragalactic IMF

Recent results have suggested that massive early-type galaxies (ETGs, i.e. ellipticals with large velocity dispersions) may have 'bottom heavy' IMFs, i.e., they contain too many low-mass stars (<1 $M_\odot$) relative to what is seen locally. This evidence has come from dynamical measurements of the mass-to-light ratio of ETGs (e.g., Treu et al. 2010; Cappellari et al. 2012) and via observations of spectral features that are dominated by low-mass, high surface-gravity stars (e.g., Conroy & van Dokkum 2012). MOS observations with an ELT would be ideal to test the latter of these, using diagnostic features in the red-optical or near-IR. This science case may be complimentary to investigations of the IMF with E-ELT near-IR imaging (MICADO). Specific features of interest include the Na I doublet (8183/8195Å) and the Wing-Ford FeH band (spanning ~9850-10200Å), and only modest spectral resolution is required (~5-8Å is sufficient) but at high S/N (~200). Three types of studies suggest themselves:

1. To quantify the radial variation of the spectral features, i.e. if the inferred IMF differs as a function of radius in galaxies. The inner regions of ETGs are thought to have formed in massive bursts of star formation (resulting in a large velocity dispersion), while the outer parts were thought to be accreted later as smaller dwarf galaxies through minor mergers. Quantifying any radial differences in the mass function will be an important constraint on galaxy assembly models (and multiple observations at fixed radii will enable summing of spectra to increase the S/N).

2. Spatially-resolved (AO-corrected) spectroscopy would enable studies down to scales of ~20-50 pc in nearby ETGs (e.g., in the Virgo Cluster). We can then construct surface-brightness fluctuation maps on small scales (e.g., Conroy et al. in prep). If the apparent variations are truly caused by differences in the IMF, and not by unknown features in bright red giants (which dominate the total light of ETGs in the optical/near-IR), then the strength of the spectral features should be equivalent in both 'bright' and 'faint' spatial pixels.

3. MOS observations combined with the sensitivity of the E-ELT will allow study of individual globular clusters and the host galaxy at the same time, using the same methods. This will test if the GCs also show IMF variations and will enable comparisons with the population of the host galaxy – conclusively testing whether globulars are drawn from the same parent populations as the stellar content of the host galaxy.

Each of these cases would provide valuable input to models of star formation, as the IMF is the key prediction which is compared to observations – if the IMF were truly demonstrated to vary, it would have a drastic impact on our understanding of galaxy evolution.

## 8.2 High-redshift AGN

Spectroscopy of large samples of high-$z$ AGN be used to investigate key issues such as:

1. Scenarios for the formation of the black hole (BH) seeds, which will eventually grow up to form the supermassive black holes (SMBHs) seen in most galaxy bulges. Two main scenarios have been proposed to date: the monolithic collapse of rare ~$10^5$ $M_\odot$ gas clouds to BHs, or an early generation of more common ~100 $M_\odot$ BHs produced by the SNe from the first (Pop III) stars. The two scenarios predict different slopes for the SMBH mass functions at $z$>6, with the lighter, seed BHs producing a steeper SMBH mass function (e.g. Tanaka et al. 2012, Lamastra et al. in prep.). Since accretion at high-z is likely to occur at, or close to, its Eddington luminosity, spectroscopic identifications of low-to-moderate luminosity AGN at $z$>6 will be vital to distinguish between the two scenarios.

2. The evolution of the AGN duty-cycle vs. redshift can help in disentangling different AGN triggering mechanisms; two main mechanisms have been suggested: galaxy encounters and internal galaxy dynamics. The first predicts an increase in the AGN duty-cycle by a factor of 10-20 from $z$~0 to $z$~4-5 followed by a sharp decrease at higher redshift. The peak at $z$=4-5 is due to the decrease of the rate of galaxy interaction above the redshift at which groups formed ($z$~4), with its decrease toward low redshift due to high relative velocities and low densities (Fiore et al. 2012). An accurate measure the AGN duty cycle at $z$>4-5 would then prove or dismiss the galaxy encounter scenario.

3. As alluded to in SC1, the AGN contribution to reionization (and to the heating of the IGM and its effect on structure formation) remains unclear. The number density of luminous QSOs is sharply peaked at $z$=2-3 and rapidly decreases toward both higher and lower redshifts; these objects do not appear to be responsible for heating the IGM exact at $z$=2-3. However, at larger redshifts the density of low-to-moderate luminosity AGN can be relatively high, and their duty cycle can reach a peak at $z$=4-5 (see previous point); these objects may contribute to hydrogen ionization at such high redshifts. Spectroscopic identification of low-to-moderate AGN at $z$>4-5 is needed to measure their luminosity function, which is crucial to assess the global AGN contribution to the high-z IGM heating.

Low-to-moderate luminosity AGN at $z$>4-5 (log $L_{bol}$ = 44-45) have faint near-IR counterparts, $H$=25-27 (Fiore et al. 2012). Spectroscopy of such faint objects is hardly feasible with present instrumentation. *JWST*-NIRSpec may reach such faint magnitudes, but the space density of high-z AGN (0.2-2 arcmin$^{-2}$) would make it difficult to collect a large statistical sample. A multiplexed ELT-MOS, with ~100 channels over a 7×7 arcmin field match neatly to the density of high-$z$ AGNs, enabling efficient spectroscopic identification of large samples. Targets can be selected from existing *Chandra/Spitzer* data (or future X-ray/IR facilities).

## 8.3 Cepheids and the Extragalactic Distance Scale

Cepheids are a well-known primary distance indicator used to calibrate the extragalactic distance scale (and, in turn, the Hubble constant $H_0$) through their Period-Luminosity (PL) relations. Several factors prevent us from obtaining an accuracy better than a few percents on $H_0$, in particular the metallicity dependence of the PL relations (Freedman et al. 2012). For PL-relations in the *V*-band no agreement has been reached between models and observations (Romaniello et al. 2008), while the metallicity dependence is probably marginal for PL-relations in the *K*-band (Bono et al. 2010).

It seems promising to use Period-Wesenheit (PW) relations instead where the Wesenheit function has been constructed to be reddening-free. The zero-point of the $PW_{VI}$ relation appears to be strongly dependent on metallicity (Storm et al. 2011) but this preliminary result is based on only six Cepheids in the SMC.

There are numerous observations of Milky Way Cepheids, but they have a limited range of metallicities and periods ([Fe/H] >-0.5 dex, P<90d). [Fe/H] goes down to –0.9 dex for Cepheids in the Magellanic Clouds, but only a few dozens of stars have direct metallicity measurements. This should improve with the next generation of MOS instruments on existing telescopes (e.g. VLT-MOONS) but there are no direct metallicity measurements for the numerous Cepheids discovered in external galaxies at greater distances. Their metallicity is therefore determined from emission lines of HII regions or planetary nebulae, assuming that they have the same metallicity at a given galactocentric distance and a linear correlation between [O/H] and [Fe/H].

Only an ELT-MOS will have the sensitivity required to determine the metallicity of individual Cepheids in each galaxy where they have been detected, particularly in the young, metal-poor systems that are mostly beyond the grasp of current and future facilities. While current methods require high-resolution in the visible, it is likely (but not firmly established yet) that the metallicity of Cepheids can also be determined from low-resolution observations in the *J*-Band, as for RSGs (see Sect. 6.4; Davies et al. 2010; Evans et al. 2011).

## 8.4 Exoplanets in Local Group galaxies

Although more than 600 exoplanets are known to exist, they are all confined to a few parsec from the Sun. We have no idea of the dependence of the environment on planet formation and evolution. Fundamentally, the question is whether exoplanets exist also in other galaxies, and whether they are similar to those found in the Milky Way? These questions can be addressed if we can measure precise radial velocities ($\Delta v \leq 10 ms^{-1}$) for giant stars in Local Group galaxies; sufficient to detect 'hot Jupiters'. The Sagittarius Dwarf and the Magellanic Clouds have suitable targets with magnitudes in the range of 18-20 mag which will require the sensitivity of the E-ELT for such observations. It will be necessary to monitor the radial velocities of many stars in such a search, hence the importance of MOS observations.

# 9 ELT-MOS: TOP-LEVEL REQUIREMENTS

Table 3 summarises the top-level requirements that flow-down from each of the science cases. As noted earlier, for the total patrol field (on the sky) from which science targets can be selected we adopt the maximum field available in the current design of the E-ELT (equivalent to a 7' diameter).

Table 3: Summary of top-level requirements from each Science Case

| Case | Target densities | FoV/target | Spatial resolution | λ-coverage (µm) | R |
|---|---|---|---|---|---|
| SC1 | 1-2 arcmin$^{-2}$ | 2" × 2"[3] | 40-90 mas | 1.0-1.8<br>*1.0-2.45* | 5,000 |
| SC1 | 10s arcmin$^{-2}$ | – | (GLAO) | 1.0-1.8<br>*1.0-2.45* | >3,000 |
| SC2 | 1-2 arcmin$^{-2}$ | 2" × 2" | 50-80 mas | 1.0-1.8<br>*1.0-2.45* | 5,000 |
| SC2 | 10s arcmin$^{-2}$ | – | (GLAO) | 1.0-1.8<br>*1.0-2.45* | > 3,000 |
| SC3 | ≥ ~20 arcmin$^{-2}$ | – | (GLAO) | 0.8-1.7 | ≥5,000<br>*~10,000* |
| SC4 | 0.5-1 arcmin$^{-2}$ | 2" × 2" | (GLAO) | 0.4-1.0<br>*0.37-1.0* | 5,000<br>*10,000* |
| SC5 | Dense | 1" × 1"<br>*1.5" × 1.5"* | ≤75 mas<br>*20-40 mas* | 1.0-1.8<br>*0.8-1.8* | 5,000 |
| SC5 | 10s arcmin$^{-2}$ | – | (GLAO) | 0.4-1.0 | ≥5,000<br>*≥10,000* |
| SC6 | 10s arcmin$^{-2}$ | – | (GLAO) | 0.41-0.46 & 0.60-0.68<br>*0.38-0.46 & 0.60-0.68* | ≥15,000<br>*≥20,000* |

Thus, in the context of the sources for MOS observations we then divide the cases into two types:

- 'High definition': Observations of tens of channels at fine spatial resolution, with MOAO providing high-performance AO for selected sub-fields in the focal plane (e.g. Rousset et al. 2010).
- 'High multiplex': Integrated-light (or coarsely resolved) observations of >100 objects at the spatial resolution delivered by GLAO.

These two types of observations entail two sets of requirements. The requirements for the IGM case (SC4 in Section 5) are slightly different as they require optical IFU observations, but without strong requirements on the spatial resolution (i.e. GLAO is sufficient). The best aperture for the single-object mode is that which maximises the signal-to-noise ratio for point-like sources in GLAO conditions; given the expected GLAO performance of the E-ELT this was shown to be ~0.9" (Navarro et al. 2010).

Of course, a conceptual design for an ELT-MOS needs to be both feasible and affordable, thus it is unrealistic to attempt to satisfy all of the 'goal' requirements simultaneously. The next stage will be to prioritise the science requirements from the different cases via trade-offs which take into account both technical and operational feasibilities. This is intended to be done by middle of 2013, also accounting for the outcome of the ESO E-ELT Instrumentation Workshop in February 2013.

## 9.1 A link to a high-resolution (HIRES) spectrograph

The two modes of MOS observations advanced here (high multiplex, high definition) imply different 'pick-offs' to select targets in the E-ELT focal plane, analogous to the single-object and IFU modes which feed the VLT-FLAMES Giraffe spectrograph. Within such architecture one could envisage a link to feed a high-resolution spectrograph (similar to that from FLAMES to UVES), which could provide observations in parallel to, e.g., high-multiplex observations, for high-resolution spectroscopy of QSO sight-lines, bright stars etc.

---

[3] Minimum size is 1"×1" if on/off sky subtraction is used.

# 10 CONCLUSION

The Universe includes hundreds of billions of galaxies, each of them being populated by hundreds of billions of stars. Astrophysics aims to understand the complexity of an almost incommensurable number of stars, stellar clusters and galaxies, including their spatial distribution, their formation and their current interactions with the interstellar and intergalactic media. A considerable fraction of discoveries in astrophysics require statistics, which can only be addressed by a MOS. A visible/near-IR MOS with capacities adapted from stellar physics to cosmology is technically feasible as recent studies have demonstrated that key issues like sky background subtraction and multi-object AO can be solved.

The E-ELT, which will be the world's largest optical/IR telescope in the 2020s, has to be equipped with a MOS that allows the largest discovery space. This White Paper highlights a number of key science cases. We must add, however, that we have been very impressed by the far larger number of science cases that have been presented during the ELT-MOS meeting in Amsterdam (October 2012) and several national ELT-MOS meetings in different European countries and in Brazil. To account for most of them, we have defined two preferential modes, one with high multiplex (HMM), and one with high spatial definition (MOAO IFUs, HDM). During the writing of the White Paper, we have realised that almost all sciences cases will highly benefit from the combination of these two modes.

In summary, the MOS at the E-ELT will be unique to probe the sources of reionisation, to investigate their physics, to study the galaxy mass-assembly history including high-$z$ dwarves, to describe the distribution of the IGM, as well as probing resolved stars at unprecedented distances, from the outskirts of the Local Group for main sequence stars, to a significant volume including nearby galaxy clusters for luminous red supergiants.

## List of Abbreviations

| | |
|---|---|
| ALMA | Atacama Large Millimetre/submillimetre Array |
| AO | Adaptive Optics |
| CaT | Calcium Triplet |
| E-ELT | European Extremely Large Telescope |
| ETG | Early-Type Galaxy |
| EW | Equivalent Width |
| GLAO | Ground-Layer Adaptive Optics |
| HST | Hubble Space Telescope |
| IFU | Integral Field Unit |
| IGM | Inter-Galactic Medium |
| IMF | Initial Mass Function |
| IR | Infrared |
| ISM | Inter-Stellar Medium |
| JWST | James Webb Space Telescope |
| LBG | Lyman-break galaxy |
| LSBG | Low-Surface-Brightness Galaxy |
| MDF | Metallicity-Distribution Function |
| MOAO | Multi-Object Adaptive Optics |
| MOS | Multi-Object Spectrograph |
| MSTO | Main-Sequence Turn-Off |
| NGS | Natural Guide Star |
| RGB | Red Giant Branch |
| RSG | Red Supergiant |
| SED | Spectral Energy Distrubtion |
| SFR | Star-Formation Rate |
| UV | Ultraviolet |
| VISTA | Visible and Infrared Survey Telescope for Astronomy |
| WMAP | Wilkinson Microwave Anisotropy Probe |